\documentclass[aps,superscriptaddress,tightenlines,nofootinbib,floatfix,longbibliography,notitlepage]{revtex4-2}
\usepackage[left=18mm,right=19mm,top=23mm,bottom=16mm]{geometry}
\usepackage{amsmath,amssymb}
\usepackage{bm}
\usepackage{comment}
\usepackage{graphicx}
\usepackage[dvipsnames]{xcolor}
\usepackage{slashed}
\usepackage[
    colorlinks=true,
    allcolors=blue
]{hyperref}

\providecommand{\repositoryInformationSetup}{} 
\repositoryInformationSetup

\usepackage{xspace}
\usepackage{bbm}

\newcommand{\repoURL}{https://github.com/evanberkowitz/latex-base}

\newcommand{\secref}[1]{Sec.~\ref{sec:#1}}
\newcommand{\Secref}[1]{Section~\ref{sec:#1}}

\newcommand{\Appref}[1]{Appendix~\ref{sec:#1}}
\newcommand{\tabref}[1]{Tab.~\ref{tab:#1}\xspace}

\newcommand{\figref}[1]{Fig.~\ref{fig:#1}\xspace}

\renewcommand{\eqref}[1]{(\ref{eq:#1})\xspace}

\newcommand{\R}[1]{Ref.~\cite{#1}}

\newcommand{\Refs}[1]{Refs.~\cite{#1}}

\newcommand{\oneover}[1]{\ensuremath{\frac{1}{#1}}}                             

\let\builtinLaTeX\LaTeX
\def\LaTeX{\builtinLaTeX\xspace}

\providecommand{\orgname}[1]{#1}
\providecommand{\orgaddress}[1]{#1}
\providecommand{\postcode}[1]{#1}
\providecommand{\city}[1]{#1}
\providecommand{\country}[1]{#1}

\newcommand{\bonn}{
	\orgname{%
		Helmholtz-Institut f\"{u}r Strahlen- und Kernphysik, 
		Rheinische Friedrich-Wilhelms-Universit\"{a}t 
	}%
	\orgaddress{%
		\postcode{53115} \city{Bonn}, \country{Germany}
	}%
}

\newcommand{\casa}{
	\orgname{%
		Center for Advanced Simulation and Analytics, 
		Forschungszentrum J\"{u}lich,
	}%
	\orgaddress{%
		\postcode{52428} \city{J\"{u}lich}, \country{Germany}
	}%
}

\newcommand{\edinburgh}{
	\orgname{%
		School of Physics and Astronomy, The University of Edinburgh
	}%
	\orgaddress{%
		Scotland, \country{UK}
	}
}

\newcommand{\ias}{
	\orgname{%
		Institute for Advanced Simulation 4, 
		Forschungszentrum J\"{u}lich,
	}%
	\orgaddress{%
		\postcode{52428} \city{J\"{u}lich}, \country{Germany}
	}%
}

\newcommand{\jsc}{
	\orgname{%
		J\"{u}lich Supercomputing Center, 
		Forschungszentrum J\"{u}lich, 
	}%
	\orgaddress{%
		\postcode{52428} \city{J\"{u}lich}, \country{Germany}
	}%
}

\newcommand{\kitp}{
	\orgname{%
		Kavli Institute for Theoretical Physics,
		University of California,
	}%
	\orgaddress{%
		\city{Santa Barbara, CA} \postcode{93106}, \country{USA}
	}%
}

\newcommand{\uvi}{
	\orgname{%
		Department of Chemical and Physical Sciences, College of Mathematics, University of the Virgin Islands,
	}%
	\orgaddress{%
		10000 Castle Burke,
		\city{Kingshill, St. Croix}, \country{US Virgin Islands} \postcode{00850}
	}
}

\begin{document}

\title{Fully ergodic simulations using radial updates}

\author{Finn L. Temmen}
\email{f.temmen@fz-juelich.de}
\affiliation{\ias}
\author{Evan Berkowitz}
\email{evan.berkowitz@uvi.edu}
\affiliation{\ias}
\affiliation{\jsc}
\affiliation{\casa}
\affiliation{\uvi}
\affiliation{\kitp}
\author{Anthony Kennedy}
\email{Tony.Kennedy@ed.ac.uk}
\affiliation{\edinburgh}
\author{Thomas Luu}
\email{t.luu@fz-juelich.de}
\affiliation{\ias}
\affiliation{\bonn}
\author{Johann Ostmeyer}
\email{ostmeyer@hiskp.uni-bonn.de}
\affiliation{\bonn}
\author{Xinhao Yu}
\email{X.Yu-34@sms.ed.ac.uk}
\affiliation{\edinburgh}

\date{\today}

\begin{abstract}
		A sensible application of the Hybrid Monte Carlo (HMC) method is often hindered by the presence of large - or even infinite - potential barriers.
These potential barriers separate the configuration space into distinct sectors and can lead to ergodicity violations that bias measurements.
	In this work, we address this problem by augmenting HMC with a multiplicative Metropolis-Hastings update in a so-called ``radial direction'' of the fields which enables crossing the potential barriers and ensures ergodicity of the sampling algorithm at comparably low computational cost.
	We demonstrate the algorithm on a simple toy model and show how it can be applied to the fermionic Hubbard model describing physics ranging from an exactly-solvable two-site system to the $C_{20}H_{12}$ perylene molecule.  
	Our numerical results show that the radial updates successfully remove ergodicity violations, while simultaneously reducing autocorrelation times.
\end{abstract}

\maketitle

\section{Introduction}\label{sec:intro}

The Hybrid Monte Carlo (HMC) method is one of the most successful tools in the simulation of lattice field theories.
However, despite its many advantages, its application often faces challenges due to manifolds of vanishing fermion determinant which result in infinite potential barriers that lead to diverging force terms in the molecular dynamics evolution.
This separates regions in configuration space and results in ergodicity violations of the algorithm.
One important example is the freezing of global topological fluctuations in lattice quantum chromodynamics (QCD) discretizations with chiral fermions~\cite{Alles:1996vn} or near the continuum limit, where the integer-valued topological charge cannot be smoothly changed by molecular dynamics.
Another prime example that exhibits configuration space partitioning is the fermionic Hubbard model in the particle/hole basis using the exponential discretization and formulated in terms of a non-compact auxiliary field by means of a continuous Hubbard-Stratonovich transformation.
In this scenario, the action diverges due to vanishing fermion determinants, giving rise to insurmountable potential barriers that trigger an ergodicity problem in HMC simulations~\cite{Ergodicity_Hubbard}.
Reliable application of HMC necessitates the development of strategies to circumvent the potential barriers; \R{Ergodicity_Hubbard} explores coarser molecular dynamics integration and discrete jumps while other approaches introduce a complexified auxiliary field over which to integrate~\cite{AlgorithmForTheSimulationOfManyElectronSystems,Revisiting_HQMC_for_Hubbard,HMCExtendedHubbardModel,ConfPhaseTransitionGraphene}.
Normalizing flows represent a more radical departure from HMC for sampling these models~\cite{FlowsForHubbard} and do not use molecular dynamics evolution to generate the proposed updates.
In this work, we propose another method, which interleaves the HMC simulation with so-called \emph{radial updates} to overcome the ergodicity problem posed by the infinite potential barriers. 
Radial updates refer to multiplicative Metropolis-Hastings updates in a radial direction of non-compact fields that enable jumps over potential barriers and thus ensure ergodicity at comparably low computational cost.

In the following sections, we commence by reviewing the HMC algorithm and discussing ergodicity violations arising from potential barriers in \Secref{HMC}.
Then, in \Secref{Radial}, we define the radial updates and apply them to a simple toy model to demonstrate key salient features.
We then apply radial updates to the Hubbard model in \Secref{Results}, and examine their ability to overcome the ergodicity violations, their scaling properties to larger system sizes, and their efficacy in a realistic simulation of the $C_{20}H_{12}$ perylene molecule.
We conclude in \Secref{Summary}.

\section{Hybrid Monte Carlo}
\label{sec:HMC}
The fundamental objective of lattice field theory simulations is the computation of expectation values of observables $\mathcal{O}$, defined by high-dimensional integrals
\begin{equation}
	\label{eq:path_int}
	\langle \mathcal{O}\rangle =Z^{-1}\int \mathcal{D}\phi~ \mathcal{O}[\phi] e^{-S[\phi]}
	\quad \text{with} \quad 
	Z= \int \mathcal{D}\phi~  e^{-S[\phi]}.
\end{equation}
This is achieved by generating a characteristic ensemble of field configurations $\{\phi^{(i)}\}_{i=1}^{N_{\mathrm{conf}}}$ and estimating expectation values of observables \eqref{path_int} using Monte Carlo integration
\begin{equation}
	\langle \mathcal{O}\rangle \approx \frac{1}{N_{\mathrm{conf}}}\sum_{i=1}^{N_{\mathrm{conf}}} \mathcal{O}[\phi^{(i)}], 
	\quad \text{where} \quad 
	\phi^{(i)} \sim e^{-S[\phi]}/Z.
\end{equation}
So, the central task amounts to sampling field configurations $\phi^{(i)}$ according to the Boltzmann distribution $p[\phi]=e^{-S[\phi]}/Z$; a task that is amenable to Markov chain Monte Carlo (MCMC) methods.

MCMC methods are a class of algorithms that successively construct a Markov chain $\{\phi^{(i)}\}_{i=1}^{N_{\mathrm{conf}}}$, where each state $\phi^{(i)}$ only depends on its most recent predecessor $\phi^{(i-1)}$.
In the Markov chain, the transition from a configuration $\phi$ to $\phi'$ is governed by a transition probability $\Omega(\phi \rightarrow \phi')$ and its distribution converges to a target distribution $p[\phi]$ if it is ergodic\footnote{Strictly speaking, the algorithm should exhibit geometric convergence, which - on a non-compact manifold - can be achieved by fulfilling sufficient conditions for Harris' ergodic theorem, namely Doeblin's condition and the strong geometric drift condition. By slight abuse of nomenclature, we will refer to this as \emph{ergodicity} throughout this work. For further details we refer the reader to \R{original_radial_update}.} and satisfies the detailed balance condition
\begin{align}
	\label{eq:detailed_balance}
	p[\phi] \Omega(\phi \rightarrow \phi') = p[\phi'] \Omega(\phi' \rightarrow \phi).
\end{align}
The method of obtaining a new state in $\phi'$ in the Markov chain depends on the specific algorithm.

A commonly used algorithm is the Hybrid Monte Carlo (HMC) method \cite{HMC} that numerically evolves field configurations through Hamiltonian dynamics.
Specifically, a set of conjugate momenta $\{\pi_i\}_{i=1}^d$ are introduced for the field $\{\phi_i\}_{i=1}^d$, and one samples the joint $(\phi, \pi)$ space according to the probability distribution with partition function
\begin{align}
	\label{eq:HMC Z}
	\mathcal{Z} = 1 \times Z = \oneover{\mathcal{N}}\int \mathcal{D}\pi\; e^{-\pi^2/2}\; Z = \oneover{\mathcal{N}} \int \mathcal{D}\pi\; \mathcal{D}\phi\; e^{-\pi^2/2 - S[\phi]}
\end{align}
for an irrelevant normalization $\mathcal{N}$ which cancels from all observables.
HMC interprets the combined exponential in terms of the artificial Hamiltonian
\begin{align}
	\label{eq:Hamiltonian}
	H[\phi, \pi] = \frac{1}{2}\sum_{i=1}^d \pi_i^2 + S[\phi]
\end{align}
where the action $S$ plays the role of the potential.
A new configuration is obtained by evolving a starting $(\phi, \pi)$ configuration over a fictitious time $\tau$ according to Hamilton's equations
\begin{align}
	\label{eq:MD}
	\frac{d\pi}{d\tau} 
	= -\frac{\partial H}{\partial \phi} 
	= -\frac{\partial S}{\partial \phi} 
	\quad\quad \text{and} \quad\quad 
	\frac{d\phi}{d\tau} 
	= \frac{\partial H}{\partial \pi} 
	= \pi.
\end{align}
Since the artificial energy and the measure are conserved along the \emph{molecular dynamics} (MD) trajectory produced by integrating Hamilton's equations \eqref{MD}, the resulting configuration has the same weight in $\mathcal{Z}$ as the starting configuration.
The momentum $\pi$ is refreshed at the beginning of each new HMC step according to the gaussian weight in $\mathcal{Z}$ \eqref{HMC Z}.

Critical to the correctness of HMC is that the combination of refreshing the momentum and MD integration must be able to reach every configuration of $\phi$ of nonzero weight.
If infinite potential barriers partition the configuration space, the MD trajectory will never reach some valid configurations, leading to an unwanted bias in observable estimates.
Notice that when $S$ diverges so too does the force $F[\phi] = -\frac{\partial S}{\partial \phi}$ \eqref{MD} and the MD trajectory is rightly repelled.
But if $S$ diverges everywhere on a $(d-1)$-dimensional manifold in the $d$-dimensional configuration space, i.e.\@ on a codimension-1 manifold, the MD evolution becomes trapped and cannot reach every configuration.
This bias is called an \emph{ergodicity problem}.

In a practical implementation of HMC, the integration of Hamilton's equations is inevitably done numerically.
To this end, a trajectory of length $T$ is discretized into $N_{\mathrm{MD}}$ equal steps of size $\epsilon = T/N_{\mathrm{MD}}$.
An initial sample $(\phi, \pi) \equiv (\phi(0), \pi(0))$ is then evolved to a new sample $(\tilde{\phi}, \tilde{\pi}) \equiv (\phi(T), \pi(T))$ using a numerical integration scheme.
While an area-preserving and reversible integrator ensures the phase space is conserved through the integration, discretization errors can introduce violations of the energy conservation.
Therefore, to retain detailed balance, the new configuration $\tilde{\phi}$ is proposed to the Markov chain and accepted according to the Metropolis test with probability,
\begin{align}
	\label{eq:HMC_acceptance}
	\alpha_{\mathrm{HMC}} = \min \left(1, e^{-\Delta H}\right),  
	\quad \text{where} \quad
	\Delta H = H[\tilde{\phi}, \tilde{\pi}] - H[\phi, \pi],
\end{align}
or rejected.
If the proposal is accepted then the new state in the Markov chain becomes $\phi'=\tilde{\phi}$, otherwise the proposal is rejected and the previous state recurs $\phi' = \phi$.
With perfect integration, the energy conservation is exact and every trajectory is accepted.

Coarsening the integration of the MD trajectory can facilitate crossing potential barriers by increasing the energy violations $\Delta H$ \eqref{HMC_acceptance}, at the cost of a lower acceptance rate.
While this strategy may suffice to resolve ergodicity violations, the acceptance rate can plummet and result in exploding computational costs as the system size grows.
More specifically, the algorithm may suffer from exponentially increasing autocorrelation times, which constitutes an ergodicity problem in practice, as a sensible simulation quickly becomes prohibitively expensive.
It is therefore crucial to monitor simulations for signs of ergodicity violations and to develop algorithms to mitigate them, motivating the present work.

\section{Radial Updates}
\label{sec:Radial}
In this section, we introduce radial updates as a complementary Metropolis-Hastings update to HMC simulations and showcase their ability to overcome ergodicity violations caused by potential barriers, using a toy model as an illustrative example.
\subsection{Formulation}
\emph{Radial updates} refer to a multiplicative Metropolis-Hastings (MH) update of a non-compact field $\phi = (\phi_1,\dots, \phi_d)$, which generates a new proposal by scaling the radius in field space
\begin{align}
	\label{eq:radius}
	R = \sqrt{\sum_{i = 1}^{d}\phi_i^2}.
\end{align}
The radial updates constitute a special case of updates found in more general multiplicative MH algorithms, examples of which include the Random Dive MH algorithm \cite{DuttaRDMH} and the Transformation-based MCMC method \cite{DuttaTMCMC}.
Rather than devising a new multiplicative MH algorithm, we propose to amend a standard HMC simulation with intermediate multiplicative radial updates.
This approach was originally introduced in \cite{original_radial_update} and optimized in \cite{OstmeyerRadialUpdates}, with the primary goal of proving and enhancing the general convergence properties of HMC on non-compact manifolds.

An additional benefit of incorporating radial updates is that it facilitates crossing potential barriers because they are agnostic to the diverging force.
This removes ergodicity violations that would otherwise challenge a standalone HMC algorithm.

Starting from an initial configuration $\phi = (\phi_1,\dots, \phi_d)$, the radial update is defined as follows:
\begin{itemize}
	\item[1. ] Draw an update variable $\gamma$ from a normal distribution $\mathcal{N}(\gamma |\mu = 0,\sigma_R^2)$ with a mean of zero and proposal standard deviation $\sigma_R$. 
	\item[2. ] Generate a new configuration $\phi'$ by multiplying the initial configuration $\phi$ with $e^\gamma$, i.e.
	\begin{align}
		\phi' =  (e^\gamma\phi_1,\dots,e^\gamma\phi_d).
	\end{align}
	This corresponds to rescaling the radius of $\phi$ to $R'=e^\gamma R$, motivating the term \emph{radial update}.
	Since $\gamma$ is drawn from a zero-centered normal distribution, $e^\gamma$ follows a log-normal distribution with a median of one, meaning that increases and decreases in the radius are equally probable.
	\item[3. ] Use the new configuration $\phi'$ in a Metropolis acceptance test with acceptance probability 
	\begin{align}
		\label{eq:radial_acc_prob}
		\alpha_{R} = \min\left(1, e^{-\Delta S + d\gamma }\right)
	\end{align}
	where $\Delta S = S[\phi']-S[\phi]$ is the change in action and $e^{d\gamma}$ represents the Jacobian of the multiplicative update. Notably, the acceptance probability only depends indirectly on the proposal standard deviation used to sample $\gamma$. 
\end{itemize}
The radial update, and therefore also the combined algorithm of HMC and radial update, satisfy the detailed balance condition \eqref{detailed_balance}.

In practice, employing radial updates requires additional design choices compared to a standalone HMC simulation.
These include selecting and tuning the standard deviation $\sigma_R$ of the radial updates, as well as the number of HMC steps and radial updates per iteration of the combined algorithm.
Building on the insights from \R{OstmeyerRadialUpdates}, the tuning and scaling of $\sigma_R$ for maximal efficiency will be examined in \Secref{Results}.

\subsection{Toy model: a first application}
\label{sec:toy_model_interlude}
\begin{figure}
	\centering
	\includegraphics[width = 0.8\textwidth]{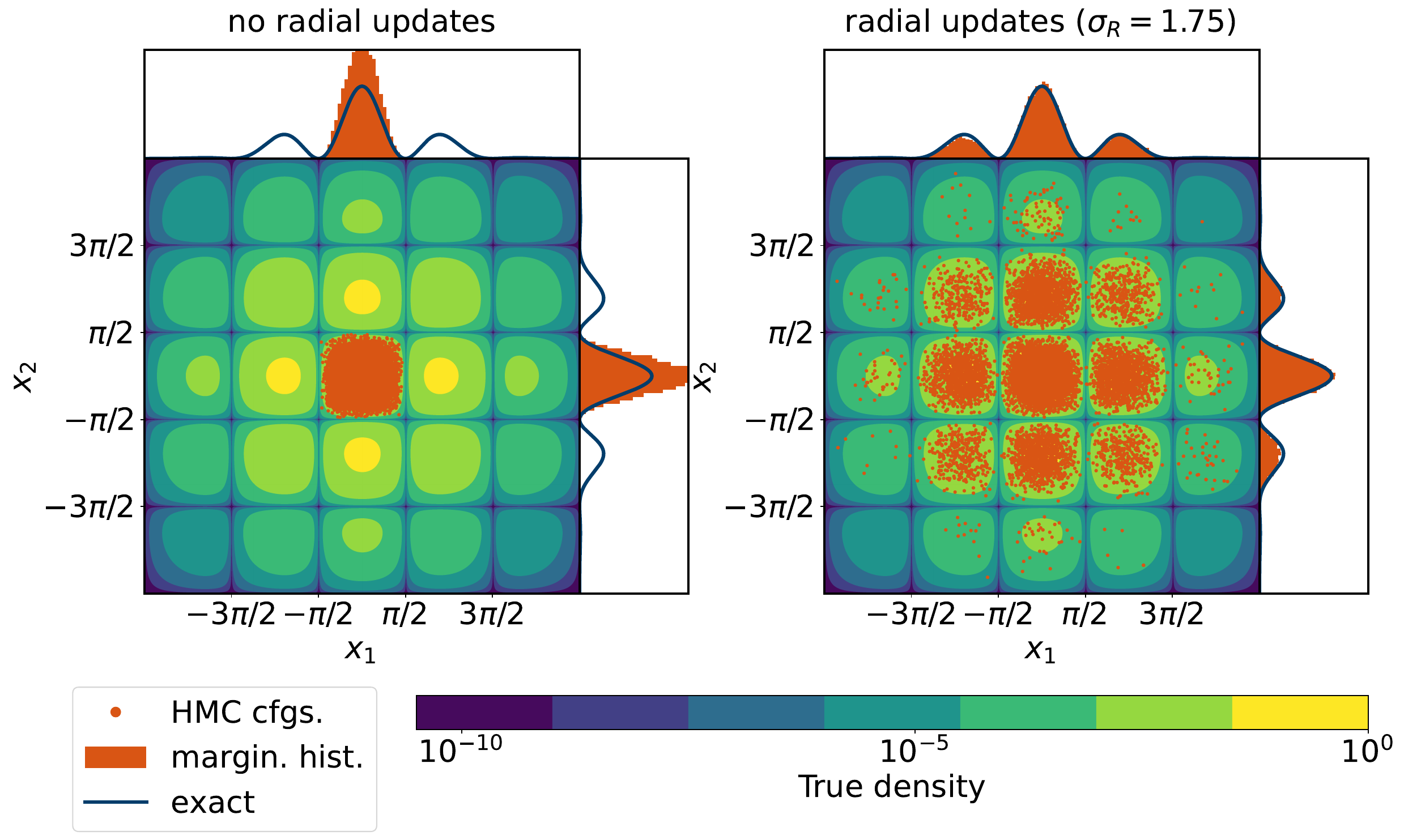}
	\caption{HMC configurations (red dots) for the toy model~\eqref{toy_model} with $d=2$ at $\beta = 0.125$.
		The configurations are compared to the exact distribution, shown as a contour plot.
		The HMC simulations were conducted  using $T=1$ and $N_{\mathrm{MD}}=12$, achieving an acceptance rate $\gtrsim 99\%$.
		The left panel shows the simulation without radial updates, while the right panel includes radial updates.
		In the latter, a single radial update was performed before each HMC step with a proposal standard deviation of $\sigma_R = 1.75$.
		Each subplot shows a total of $10^4$ configurations, obtained by recording every configuration. 
		The plots on the margins compare the marginal distribution to the histograms obtained from the visualized trajectories.}
	\label{fig:Toy_scatter}
\end{figure}

In order to demonstrate how radial updates can resolve the ergodicity violations caused by infinite potential barriers, we consider the toy model defined by the probability distribution
\begin{equation}
	\label{eq:toy_model}
	p[x]\propto \prod_{i=1}^d \cos^2(x_i)e^{-\beta x_i^2}, \quad \text{where}\quad x\in\mathbb{R}^d. 
\end{equation}
Even though this model exhibits infinite potential barriers at $x_i = \frac{2k+1}{2}\pi$ for $k\in \mathbb{Z}$, the high-dimensional path integral \eqref{path_int} for this model factorizes and is therefore easily solvable.
These barriers pose a potential challenge for solutions obtained through HMC, however, and this can be demonstrated by performing a standalone HMC simulation with fine MD integration and comparing to the exact distribution \eqref{toy_model}.
Specifically, we simulate the toy model with dimensionality $d=2$ at $\beta = 0.125$, setting $N_{\mathrm{MD}}=12$ and $T=1$, which results in an acceptance rate of $\gtrsim 99\%$.
In total, we record $10^4$ configurations, saving after each HMC step.
The resulting configurations are displayed in the left panel of \figref{Toy_scatter} and compared to the exact distribution given as a contour plot. Furthermore, histograms of the single components are compared to the marginalized distribution at the margins of the plot.
It is apparent that HMC is trapped in the center region of configuration space, which is separated from adjacent regions by the infinite potential barriers at $x_i \in \{-\pi/2,\pi/2\}$. 
Therefore, the algorithm fails to explore regions of high probability, resulting in a severe ergodicity problem.
In this example, the middle peak in the marginal distribution of the respective single components is heavily oversampled, which trivially leads to a bias in measurements of observables. 

Next we perform the same simulation but precede each HMC step with a radial update with proposal standard deviation $\sigma_R=1.75$.
The resulting configurations are displayed in the right panel of \figref{Toy_scatter} and we observe that the combined algorithm is now able to transition through the potential barriers. 
This leads to an extensive exploration of the entire configuration space and, as depicted at the margins of the plot, the simulation correctly reproduces the marginal distributions.

This concludes our initial demonstration of the radial updates' ability to overcome ergodicity violations due to infinite potential barriers.
In the following, we extend our investigations to a non-trivial model, namely the Hubbard model formulated on a non-compact space by means of a continuous Hubbard-Stratonovich transformation.
Specifically, we study the tuning of $\sigma_R$ and the overall efficacy of radial updates when increasing the number of lattice sites and spacetime dimensions.
We performed a similar scaling analysis for the toy model introduced in the present section, and the main results are summarized in \Appref{toy_model}.

\section{Hubbard model}
\label{sec:Results}
In this section, we apply the combined algorithm of HMC and radial updates, as introduced in \Secref{Radial}, to simulate the Hubbard model and explore its efficiency and scaling properties.
We begin by formulating the Hubbard model in \Secref{Hubbard}, followed by a discussion of the emergence of infinite potential barriers and the related ergodicity violations.
Next, we analyze the exactly solvable two-site model, highlighting the ergodicity problem and demonstrating the radial updates' ability to resolve it.
We then increase the system's dimensionality by increasing the number of time slices and examine the scaling properties of the proposal standard deviation and autocorrelation times.
Finally, we apply the radial updates to simulate the $C_{20}H_{12}$ perylene molecule, assessing their effectiveness in simulating a realistic system size with reasonably tuned MD integration.

The simulations and data analyses were performed using the Nanosystem Simulation Library \cite{NSL}, the comp-avg tool \cite{comp-avg}, and the analysis code provided in \R{RadialUpdatesCode}.

\subsection{Formulation}
\label{sec:Hubbard}
The Hubbard model is commonly used in condensed matter physics to describe the behavior of strongly-correlated electrons on a lattice.
We use the formulation of the Hubbard model in the so-called particle/hole basis, where the Hubbard Hamiltonian on a spatial lattice with $N_x$ sites is defined as
\begin{align}
	\label{eq:Hubbard_Hamiltonian}
	H 
	= H_K + H_U 
	= -\kappa \sum_{\langle x, y\rangle} \left(
	a_x^\dagger a_y^{\phantom{\dagger}}
	- b_x^\dagger b_y^{\phantom{\dagger}}
	\right)
	+ \frac{U}{2} \sum_x \left(
	a_{x}^\dagger a_x^{\phantom{\dagger}} 
	- b_x^\dagger b_x^{\phantom{\dagger}}
	\right)^2.
\end{align}
The Hamiltonian includes a nearest neighbor hopping term with the hopping parameter $\kappa$, along with an on-site interaction characterized by the interaction strength $U$.
The fermionic operator $a_x^\dagger,$ ($a_x^{\phantom{\dagger}}$) creates (annihilates) a spin-$\uparrow$ electron-particle at the lattice site $x$.
In contrast, the operator $b_x^\dagger,$ ($b_x^{\phantom{\dagger}}$) creates (annihilates) a spin-$\downarrow$ electron-hole at the lattice site $x$.
The partition function and expectation value of an observable $\mathcal{O}$ are given by the thermal traces
\begin{align}
	\label{eq:therm_trace}
	Z 
	= \mathrm{tr}\left(
	e^{-\beta (H_K + H_U)}
	\right) 
	\quad \text{and} \quad 
	\langle \mathcal{O} \rangle 
	= \mathrm{tr}\left( 
	\mathcal{O}e^{-\beta (H_K + H_U)}
	\right)Z^{-1}\ ,
\end{align}
respectively.
In order to facilitate the application of the HMC algorithm introduced in \Secref{HMC}, this framework has to be cast into a lattice field theory by converting the partition function and thermal expectation values \eqref{therm_trace} into path integrals \eqref{path_int}.
In the following we will only sketch the derivation of this formalism. 
See, for example, \Refs{Negele:1988vy, Brower:2011nqf, Smith:2014tha, Luu:2015gpl, RodekampPhD} for more details of this formalism.

As a first step towards a path integral formulation, the inverse temperature $\beta$ is interpreted as an extent in Euclidean time and is discretized into $N_t$ time slices with temporal lattice spacing $\Delta_t = \frac{\beta}{N_t}$.
Next, a second order Suzuki-Trotter decomposition is performed, which introduces an overall error of $\mathcal{O}(\Delta_t^2)$ and necessitates taking the continuum limit $N_t \rightarrow \infty$ to recover the exact expression.

This manipulation allows us to remove the many body interactions in $H_U$ by applying the continuous Hubbard-Stratonovich transformation
\begin{align}
	\exp\left\{ 
	-\frac{\Delta_t U}{2}\sum_x\left(
	a_x^\dagger a_x^{\vphantom{\dagger}}
	- b_x^\dagger b_x^{\vphantom{\dagger}}
	\right)^2
	\right\} 
	\propto 
	\int_{-\infty}^{\infty} \left(\prod_{x = 1}^{N_s}d\phi_{tx}\right)
	\exp\left\{
	- \frac{1}{2\Delta_t U}\sum_x\phi_{tx}^2 
	- i\sum_x\phi_{tx}\left(
	a_x^\dagger a_x^{\vphantom{\dagger}}
	- b_x^\dagger b_x^{\vphantom{\dagger}}
	\right) 
	\right\}, 
\end{align}
on every time slice $t$.
This transformation is exact up to an irrelevant overall constant and eliminates the four-fermion terms but comes at the cost of an integral over the non-compact auxiliary field $\phi$.

At this point, one can evaluate the Hilbert-space trace using fermionic coherent states, allowing the terms quadratic in fermionic creation and annihilation operators to be replaced by bilinear expressions in Grassmann variables.
Performing the resulting gaussian Grassmann integral yields the effective Hubbard action
\begin{align}
	\label{eq:Hubbard_action}
	S[\phi] 
	= \frac{1}{2U\Delta_t}\sum_{t,x} \phi_{tx}^2 
	- \log \left( \det M[\phi|\kappa]\det M[-\phi|-\kappa]\right)
\end{align}
with the fermion matrix
\begin{equation}
	\label{eq:fermion_matrix}
	M[\phi|\kappa]_{tx,t'y} = 
	\delta_{t,t'}\delta_{x,y} 
	- \left(e^{\kappa h}\right)_{xy} e^{i\phi_{tx}}\mathcal{B}_{t'}\delta_{t',t+1}.
\end{equation}
Here, $h = \Delta_t \delta_{\langle z, z'\rangle}$ is the hopping matrix and $\mathcal{B}_{t}$ encodes the anti-periodic boundary conditions with $\mathcal{B}_{t}=+1$ for $0<t'<N_t$ and $\mathcal{B}_{0}=-1$.
This \emph{exponential discretization} of the Hubbard model will serve as the basis for the analysis carried out in the remainder of this work.
Other discretizations exist, but the exponential discretization is attractive because it preserves an exact chiral symmetry of the Hubbard Hamiltonian \eqref{Hubbard_Hamiltonian}.
However, that same symmetry implies that the fermion determinant $\det \left(M[\phi|\kappa] M[-\phi|-\kappa]\right)$ has manifolds of zero weight that divide the configuration space into distinct sectors \cite{Ergodicity_Hubbard}.
As discussed in \Secref{HMC}, this makes HMC susceptible to ergodicity problems caused by infinite potential barriers that need to be addressed to ensure correct sampling.

\subsection{Two-site model: restoring ergodicity}
\label{sec:R2S_ergodicity}
We begin by considering a two-site model on a single time slice, so that a configuration of $\phi$ is given by the pair $(\phi_1, \phi_2)$.
The extremely low dimension allows for a direct visualization of the ergodicity problems posed by the infinite potential barriers discussed in \Secref{Hubbard}.
We perform simulations of the model with $U = 18$, $\kappa = 1$ and $\beta = 1$, using both the standalone HMC and HMC augmented with radial updates.
In all subsequent simulations of the Hubbard model, the MD integration is carried out using the Leapfrog integrator with the trajectory length set to
\begin{align}
	\label{eq:optT}
	T=\frac{\pi}{2}\sqrt{U\beta/N_t}, 
\end{align}
which eliminates autocorrelations arising from the harmonic part of the action \cite{FourierAcceleration}.
The number of MD steps is tuned to obtain a fine integrator with acceptance rate $>99\%$, leading to $N_{\mathrm{MD}}=60$ for $N_x=2$, $N_t=1$.
We begin by generating $10^4$ configurations, saving the configuration after each HMC step.
In the simulation with radial updates, we additionally employ one radial update per HMC step with standard deviation $\sigma_R = 1.8$.

\begin{figure}
	\centering
	\includegraphics[width = 0.8\textwidth]{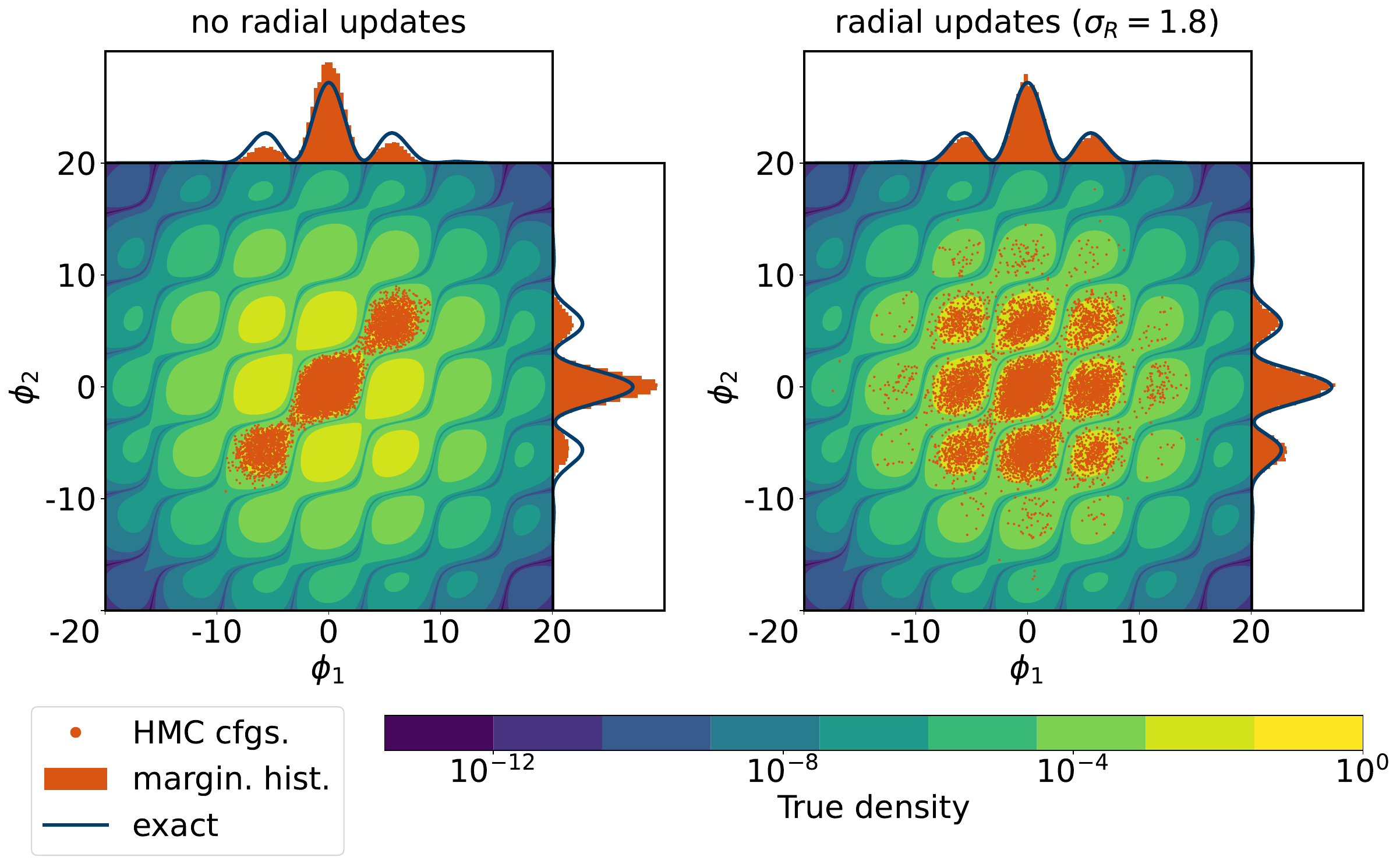}
	\caption{HMC configurations (red dots) for the two-site model on a single time slice ($N_t=1$) for $U = 18$, $\beta = 1$ and $\kappa = 1$.
		The configurations are compared to the exact distribution, shown as a contour plot.
		The HMC simulations were conducted using \eqref{optT} and $N_{\mathrm{MD}}=60$, achieving an acceptance rate $>99\%$.
		The left panel shows the simulation without radial updates, while the right panel includes radial updates.
		In the latter, a single radial update was performed before each HMC step with a proposal standard deviation of $\sigma_R = 1.8$.
		Each panel shows $10^4$ recorded configurations, with measurements taken after each HMC step.
		The plots at the margins compare the exact marginal distribution to the histograms obtained from the visualized trajectories.}
	\label{fig:Nt1_scatter}
\end{figure}

The first $10^4$ configurations in the Markov chain are visualized in \figref{Nt1_scatter} and compared to the exact probability distribution (see Sec. III.C of \R{Ergodicity_Hubbard}).
Without radial updates, shown in the left panel, HMC is confined to the middle diagonal band in configuration space, which is separated from adjacent bands by infinite potential barriers.
Consequently, the simulation is unable to sample the regions of high probability in the adjacent bands, resulting in a severe ergodicity problem.
In contrast, when radial updates are activated, as shown in the right panel of \figref{Nt1_scatter}, they facilitate transitions through the potential barriers, thereby completely resolving the ergodicity violations.
This qualitative observation is further underscored by comparing histograms of the recorded configurations to the exact marginal distributions, as presented at the margins of the two plots in \figref{Nt1_scatter}.
Without radial updates, the simulation is biased and the middle peak at $\phi_i=0$ is heavily oversampled.
However, when radial updates are applied, the simulation closely matches the exact distribution.

\begin{figure}
	\centering
	\includegraphics[width = 0.8\textwidth]{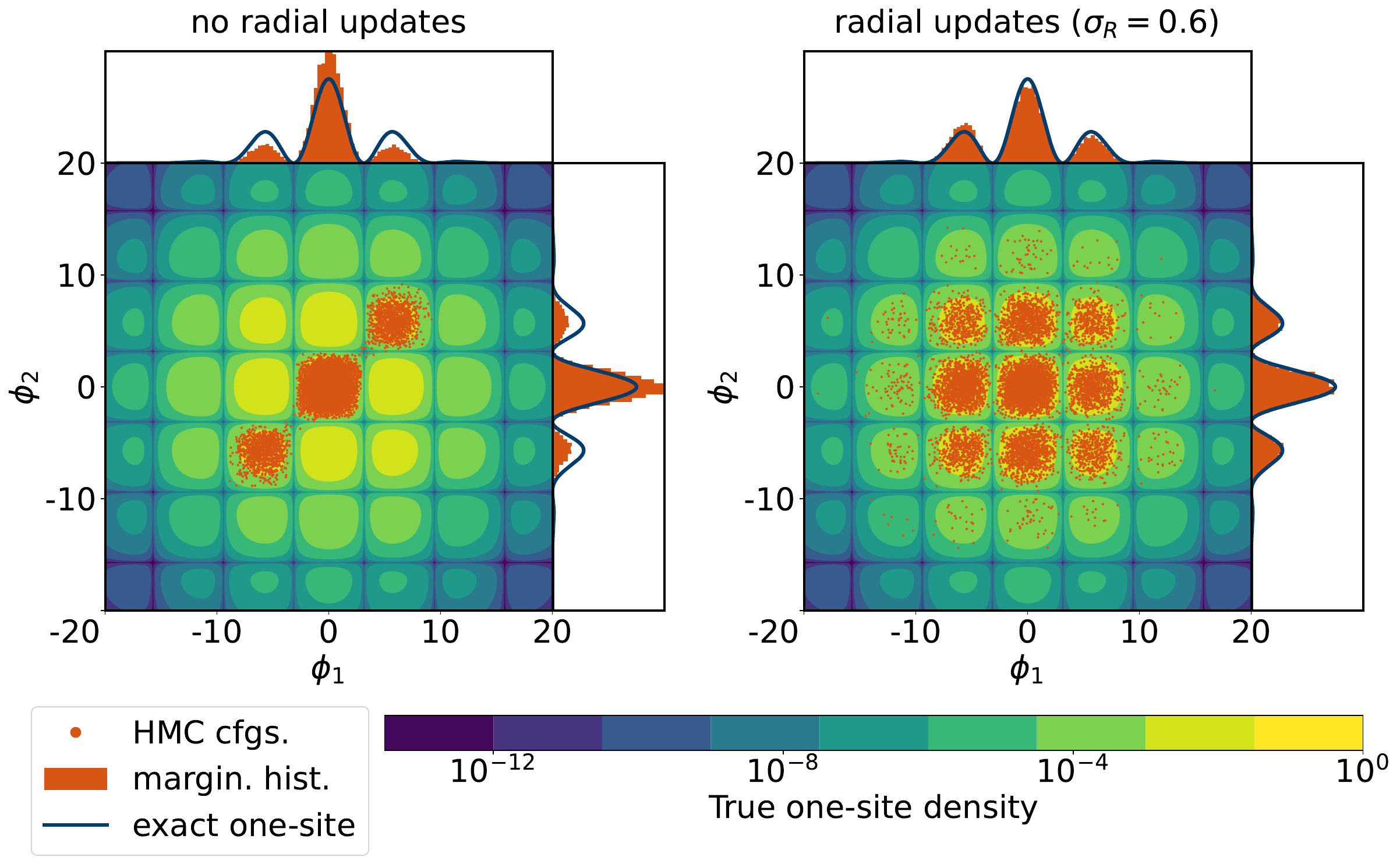}
	\caption{HMC configurations (red dots) for the two-site model with $N_t=8$ for $U = 18$, $\beta = 1$ and $\kappa = 1$.
		The configurations $\Phi=(\Phi_1, \Phi_2)$ with $\Phi_x = \sum_t \phi_{tx}$ are compared to the exact one-site distribution, recovered in the strong-coupling limit $U/\kappa \gg 1$ and shown as a contour plot.
		The HMC simulations were conducted  using an optimal trajectory length \eqref{optT} and $N_{\mathrm{MD}}=60$, achieving an acceptance rate $>99\%$.
		The left panel shows the simulation without radial updates, while the right panel includes radial updates.
		In the latter, a single radial update was performed before each HMC step with a proposal standard deviation of $\sigma_R = 0.6$.
		Each panel shows $10^4$ recorded configurations, with measurements taken after every 10th HMC step.
		The plots on the margins compare the marginal one-site distribution to the histograms obtained from the visualized trajectories.}
	\label{fig:Nt8_scatter}
\end{figure}

Next, we increase the dimensionality $d$ of the system by increasing the number of time slices $N_t$, while keeping the parameters $U$, $\beta$, and $\kappa$ unchanged.
The trajectory length $T$ is adjusted to reduce autocorrelations \eqref{optT} and $N_{\mathrm{MD}}$ is tuned to maintain a fine integrator with an acceptance rate $>99\%$.
The specific parameter choices and further details on the simulations are given in \Appref{sim_details_R2S}.
In this case, the potential barriers can still be visualized by considering $\Phi_x=\sum_t \phi_{tx}$, because in the strong-coupling limit $U/\kappa \gg 1$, the probability weights of a configuration $\Phi = (\Phi_1, \Phi_2)$ are well-approximated by the exact one-site distribution derived in \R{Ergodicity_Hubbard}.
We begin by examining the example case of $N_t = 8$, visualizing $10^4$ recorded configurations, both with and without radial updates.
In both simulations, we recorded a configuration only after every 10th step, such that the shown trajectory corresponds to a total of $10^5$ iterations of the respective algorithm.
In the simulation with radial updates we employ one radial update with standard deviation $\sigma_R = 0.6$ per HMC step.
The results are shown in \figref{Nt8_scatter}.
Similarly to the $N_t=1$ case, we observe that without radial updates, the trajectory remains confined to the middle diagonal band in the two-dimensional $\Phi$-plane.
However, with radial updates, the algorithm efficiently explores the entire configuration space, effectively restoring ergodicity.

\begin{figure}
	\centering
	\includegraphics[width = 0.6\textwidth]{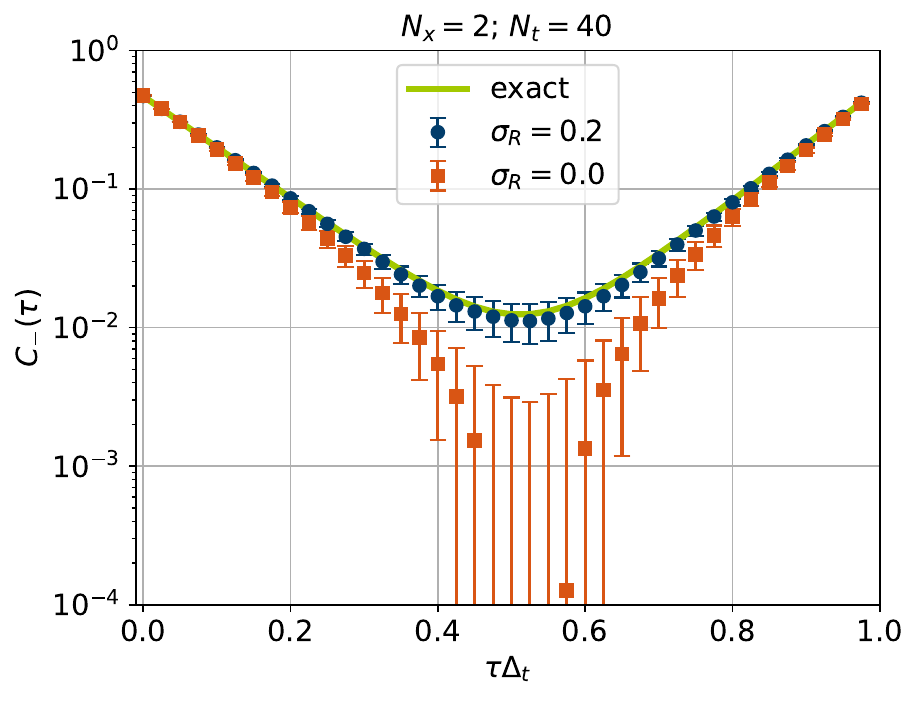}
	\caption{
		The $C_-(\tau)$ correlator of the two-site model with temporal extent $\beta=1$ and $U=18$ (see~\cite{Ergodicity_Hubbard} for a description of this correlator).  Calculations were done without radial updates (red squares) and with radial updates (blue circles), with $\sigma_R$ given in the legend.
		The exact result is given by the green solid line.
		Only simulations with radial updates agree with the exact result.
	}
	\label{fig:R2S_correlators}
\end{figure}

These ergodicity violations lead to biased measurements that, while being difficult to detect directly from simple observables, significantly impact the correctness of results.
To demonstrate this in a controlled scenario, we consider spatial correlators,
\begin{align}
	\label{eq:single particle correlator}
	C_{xy}(\tau) 
	= \left\langle a_x^{\vphantom{\dagger}}(\tau) a_{y}^\dagger (0)\right\rangle 
	= \biggl\langle \sum_t \left(M^{-1}\right)_{tx,(t+\tau) y}^{\vphantom{\dagger}} \biggr\rangle, 
\end{align}
that are projected to particular irreducible representations (irreps) $\lambda$ intrinsic to the symmetry group $\Gamma$ of the system under study,
\begin{align}
	\label{eq:R2S_correlator}
	C_\lambda(\tau) 
	= \sum_{x,y} \mathcal{A}^{\lambda}_x C_{xy}(\tau)\mathcal{A}^{\lambda\ast}_y. 
\end{align}
With translational invariance the irreps are labeled by a definite momentum $k$ and the coefficients are the Fourier components, $A^k_x \propto e^{ik\cdot x}$ up to normalization.
For the two-site model there are two possible correlators \eqref{R2S_correlator} which we can separate based on the parity of the system,
\begin{align}
	\label{eq:two-site parity projection}
	\mathcal{A}^+ &= \frac{1}{\sqrt{2}}\begin{pmatrix} 1 \\ 1 \end{pmatrix}
	&
	\mathcal{A}^- &= \frac{1}{\sqrt{2}}\begin{pmatrix} 1 \\ -1 \end{pmatrix};
\end{align}
we concentrate on the $C_-(\tau)$ correlator with $N_t = 40$ time slices and perform HMC simulations without radial updates ($\sigma_R = 0$) and with radial updates ($\sigma_R = 0.2$).
The result is depicted in \figref{R2S_correlators} and it is observed that the standalone HMC simulation significantly deviates from the exact solution, obtained from direct diagonalization using \cite{beehive}.
Meanwhile, in the simulation with radial updates, ergodicity was restored successfully and the measured correlator matches the exact result.

\subsection{Two-site model: reducing autocorrelation times}
\label{sec:R2S_tint}
We proceed with our analysis of the two-site model by investigating the effect of radial updates on autocorrelations and their scaling properties when going to larger system sizes.
Specifically, we utilize the integrated autocorrelation time to optimize the proposal standard deviation $\sigma_R$ and quantify its leading-order scaling.

The integrated autocorrelation time is an observable dependent quantity that measures the extent of correlations within a given time series.
It is defined by
\begin{equation}
	\label{eq:tint}
	\tau_{\mathrm{int},\mathcal{O}} 
	= \frac{1}{2} \sum_{t=-\infty}^\infty \frac{\Gamma_\mathcal{O}(t)}{\Gamma_\mathcal{O}(0)},
\end{equation}
in terms of the autocorrelation function
\begin{equation}
	\label{eq:autocorr_function}
	\Gamma_\mathcal{O}(t) = \bigl\langle \bigl[\mathcal{O}^{(i)} - \langle \mathcal{O} \rangle\bigr] \bigl[\mathcal{O}^{(i+t)} - \langle \mathcal{O} \rangle\bigr]\bigr\rangle, 
\end{equation}
where, in the context of lattice field theory, the $\mathcal{O}^{(i)}$ refer to subsequent measurements of the observable $\mathcal{O}$ in a Markov chain.
In practice, computing the autocorrelation time is naturally limited by a finite sample size and thus presupposes truncating the infinite sum \eqref{tint}.
Several strategies for reliably choosing the summation window have been developed and throughout this work, unless stated otherwise, we employ the method proposed in \R{WolffAutocorr}.
For a comprehensive and more detailed introduction to the estimation of errors and autocorrelations we refer the reader to \Refs{WolffAutocorr,Sokal}.
For simplicity, we will omit the subscript $\mathcal{O}$ in $\tau_{\mathrm{int},\mathcal{O}}$, unless its inclusion is necessary for clarity.

We begin by mapping out the integrated autocorrelation time $\tau_{\mathrm{int}}$ in the two-site model as a function of the proposal standard deviation $\sigma_R$.
To this end, we perform simulations across a range of proposal standard deviations $\sigma_R$ and for increasing $N_t$.
The simulations are detailed in \Appref{sim_details_R2S}.
During our analysis we considered several observables and, for clarity, we will focus on two representative observables in the following analysis.
The observables we consider are the heuristically motivated $\Phi$-plane radius, defined by
\begin{align}
	\label{eq:Phi_radius}
	\mathcal{O}_\Phi = \sqrt{\sum_x\left(\sum_t \phi_{tx}\right)^2},  
\end{align}
and $C_-(\tau)$, the single particle correlator \eqref{R2S_correlator} projected to the negative-parity irrep \eqref{two-site parity projection}. 
Technically, this correlator consists of $N_t$ components, for which we aim to reduce $\tau_{\mathrm{int}}$ to a single value for the subsequent analysis.
To achieve this, we define it as the ensemble maximum over $\tau=0,1,\dots, N_t-1$ for the component $C(\tau)$, i.e.
\begin{align}
	\label{eq:tint_correlator}
	\tau_{\mathrm{int},C}=\max_\tau \tau_{\mathrm{int},C(\tau)}.
\end{align}

\begin{figure}
	\centering
	\includegraphics[width = 0.8\textwidth]{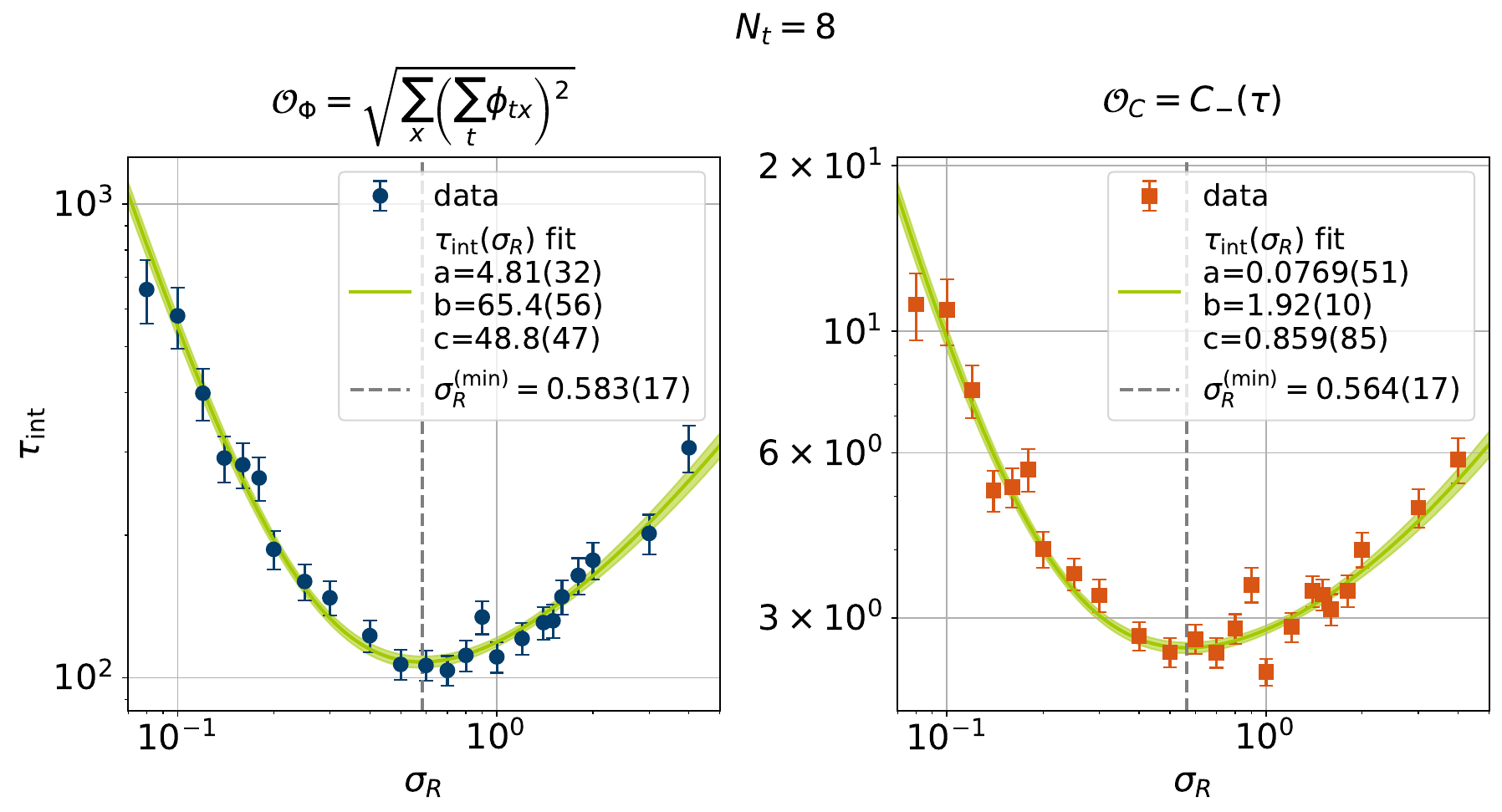}
	\caption{Integrated autocorrelation time $\tau_{\mathrm{int}}$ as a function of the proposal standard deviation $\sigma_R$ for the (left) observable $\mathcal{O}_\Phi$ \eqref{Phi_radius} and the (right) single particle correlator $\mathcal{O}_C$ \eqref{R2S_correlator}. 
		The fits (green line) are obtained using the ansatz \eqref{tint_fit}, with fit results shown in the corresponding legends.
		These results are used to estimate the optimal proposal standard deviation $\sigma_R^{\mathrm{(min)}}$ (grey dashed line), whose value is also provided in the legends.
		The underlying model is the two-site model with $N_t=8$ for $U = 18$, $\beta = 1$, and $\kappa = 1$.
		The HMC simulations were conducted using an optimal trajectory length \eqref{optT} and $N_{\mathrm{MD}}=50$, achieving an acceptance rate $>99\%$.
		Each HMC step was preceded by a single radial update and a total of $N_{\mathrm{conf}}=3\times 10^5$ configurations were recorded, with measurements taken after each HMC step.}
	\label{fig:R2S_Nt8_tint}
\end{figure}

Results for an example simulation at $N_t = 8$ are depicted in \figref{R2S_Nt8_tint}, where we observe that, for both observables, the integrated autocorrelation time initially decreases as $\sigma_R$ increases, reaches a minimum, and then starts to rise.
To describe the dependence of the integrated autocorrelation time on the proposal standard deviation $\sigma_R$, we adopt the fitting ansatz
\begin{align}
	\label{eq:tint_fit}
	\tau_{\mathrm{int}}(\sigma_R^{\vphantom{-2}}) = a\sigma_R^{-2} + b + c\sigma_R^{\vphantom{-2}}.
\end{align}
In this expression, the first term accounts for the expected behavior at small $\sigma_R$, where the proposed step size is small and almost all proposals are accepted.
This results in a random walk and a diffusive regime, where $\tau_{\mathrm{int}}\propto \sigma_R^{-2}$. 
Additionally, under the assumption of a perfect MD integration, it also captures the diverging autocorrelation time due to ergodicity violations when no radial updates are employed ($\sigma_R = 0$).
The third term quantifies the large $\sigma_R$ regime where proposed steps are large and significantly decorrelate subsequent samples.
However, as $\sigma_R$ is increased further, the acceptance rate decreases and the incremental decorrelation of accepted proposals is negligible.
Therefore, $\tau_{\mathrm{int}}$ increases with decreasing acceptance rate, resulting in the linear regime $\tau_{\mathrm{int}}\propto \sigma_R$.

\begin{figure}
	\centering
	\includegraphics[width = 0.9\textwidth]{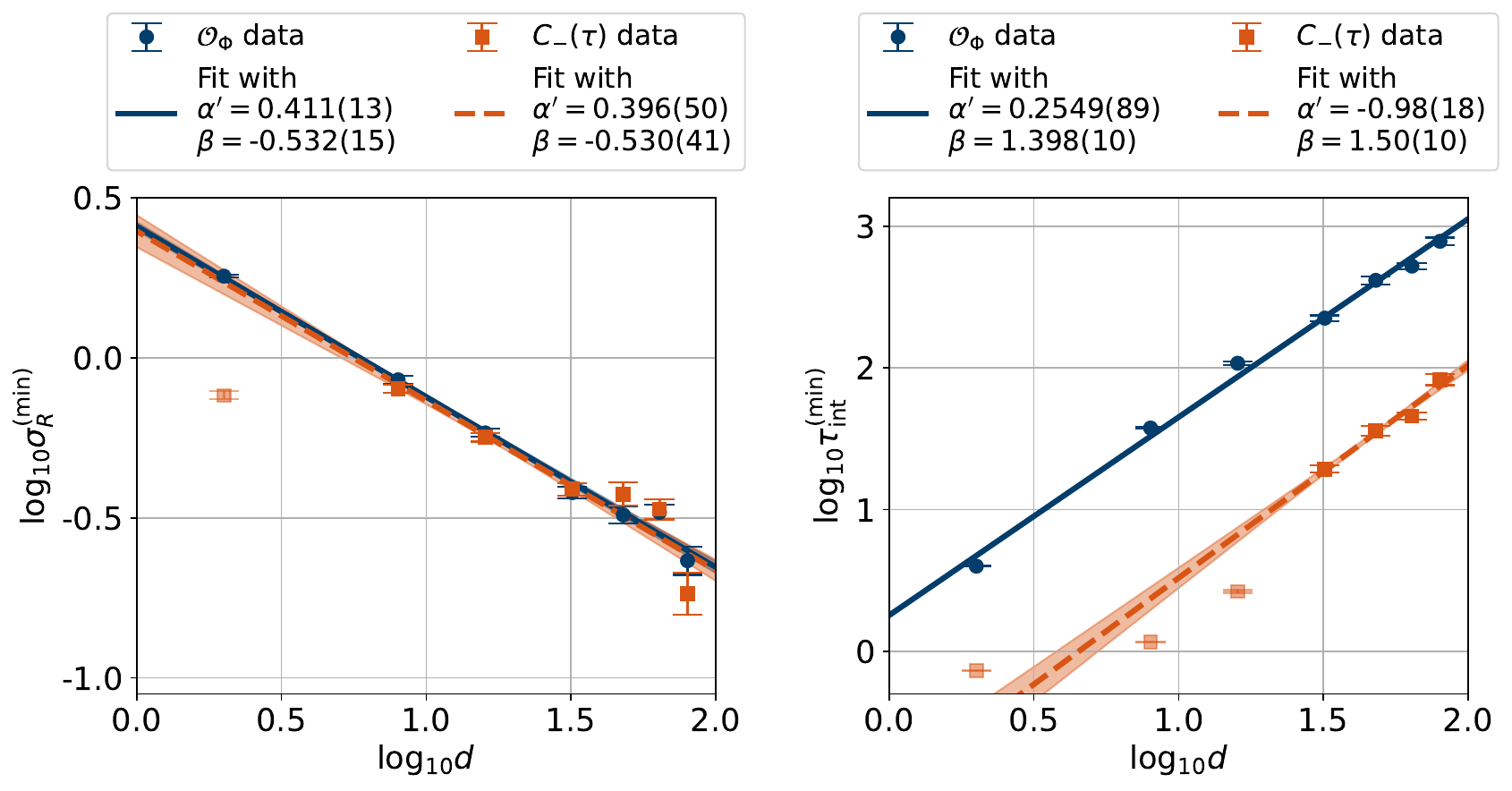}
	\caption{The left panel shows a double logarithmic plot of the position of the minimum $\sigma_R^{(\mathrm{min})}$ as a function of dimensionality $d=N_xN_t$.
		The right panel illustrates a double logarithmic plot of the integrated autocorrelation time at the respective minimum, denoted by $\tau_{\mathrm{int}}^{(\mathrm{min})}$, also as a function of the dimensionality $d$.
		The data in both panels is fitted to the respective ansatz, \eqref{sigmin_fit} and \eqref{tintmin_fit}, to determine the leading-order scaling.
		The results were obtained using $\mathcal{O}_\Phi$ \eqref{Phi_radius} shown as blue circles, and the single particle correlator $\mathcal{O}_C$ \eqref{R2S_correlator} projected to the negative-parity irrep \eqref{two-site parity projection} shown as red squares.
		Data points shown in low opacity have been excluded from the fitting procedure and fit results are provided in the legends, where $\alpha'=\log_{\mathrm{10}}\alpha$.
		The underlying model is the two-site Hubbard model with varying $N_t$ for $U = 18$, $\beta = 1$, and $\kappa = 1$.
	}
	\label{fig:R2S_scaling}
\end{figure}

The extrapolation is displayed in \figref{R2S_Nt8_tint} and throughout this work, fit results are obtained by fitting the respective ansatz to $N_{\mathrm{boot}}=10^3$ bootstrap samples of the measured data.
It is evident that the chosen ansatz effectively captures the behavior of the data and allows for a sensible estimation of the optimal proposal standard deviation, given by the position of the minimum $\sigma_R^{(\mathrm{min})}$ (dashed vertical line in \figref{R2S_Nt8_tint}).
Furthermore, by repeating this analysis for all values of $N_t$, it enables us to study the scaling behavior of the optimal proposal standard deviation with increasing dimensionality $d=N_xN_t$ which is crucial for the sensible application in a realistic simulation.
The respective estimates for $\sigma_R^{(\mathrm{min})}$ are shown in the left panel of \figref{R2S_scaling}.
Theoretical considerations suggest that the optimal proposal standard deviation should scale as $\sigma_R^{(\mathrm{min})}(d)\propto \frac{1}{\sqrt{d}} + \mathcal{O}(d^{-1})$ at leading order \cite{OstmeyerRadialUpdates}, motivating the fit model 
\begin{align}
	\label{eq:sigmin_fit}
	\sigma_R^{(\mathrm{min})}(d) = \alpha d^\beta.
\end{align}
By performing the fits, we find leading order exponents $\beta^{(\mathcal{O}_\Phi)}=-0.532(15)$ and $\beta^{(\mathcal{O}_C)}=-0.530(41)$, which closely match the predicted $d^{-0.5}$ scaling.
Additionally, we use the fit model \eqref{tint_fit} to estimate the integrated autocorrelation time at the minimum, denoted by $\tau_{\mathrm{int}}^{(\mathrm{min})}$. 
The results are shown in the right panel of \figref{R2S_scaling} and, like the sampling width \eqref{sigmin_fit}, we adopt the fit ansatz 
\begin{align}
	\label{eq:tintmin_fit}
	\tau_{\mathrm{int}}^{(\mathrm{min})}(d) = \alpha d^\beta
\end{align}
to determine the leading-order scaling as a function of dimensionality.
The resulting fits highlight that, when employing radial updates, the integrated autocorrelation time scales polynomially and thus does not exhibit signs of ergodicity violations anymore.

Moreover, it is worth noting that across all chosen values of $N_t$, we find a radial acceptance rate of approximately $30 \%$ close to the respective optimal values for the proposal standard deviation $\sigma_R^{(\mathrm{min})}$.
This empirical finding suggests that the standard deviation could also be fine-tuned by aiming for a specific range of radial acceptance rate and we will discuss this approach more in \Secref{perylene}.

In this section, we demonstrated the radial updates' ability to restore ergodicity in the two-site model while substantially reducing autocorrelation times.
Naturally, the aim of this study is to examine the efficacy of radial updates in realistic simulations, which is explored in the next section using the $C_{20}H_{12}$ perylene molecule as an example.
In this context, performing a realistic simulation entails two central aspects.
First, we have to scale up the system's size in terms of spatial lattice sites and the number of spatial dimensions, ultimately approaching the complexity of molecules and materials of interest.
As an intermediate step toward this goal, we have also investigated the Hubbard model on a $2\times 2$ square lattice and the results, detailed in \Appref{Sq4S}, support the findings of the present section.
Second, the simulations discussed in this section utilized a very fine MD integrator to highlight ergodicity violations caused by infinite potential barriers.
However, a fine MD integration requires a large number of MD steps, making it computationally expensive.
A more common approach is tuning $N_{\mathrm{MD}}$ to achieve an acceptance rate of approximately $70\%$, which not only renders the simulation more efficient but also enables tunneling through potential barriers due to small energy violations.
The interplay of coarse MD integration and radial updates is further quantified in the scaling analysis of the toy model \eqref{toy_model}, which is detailed in \Appref{toy_model}.

\subsection{Perylene}
\label{sec:perylene}
\begin{figure}
	\centering
	\includegraphics[width = 0.9\textwidth]{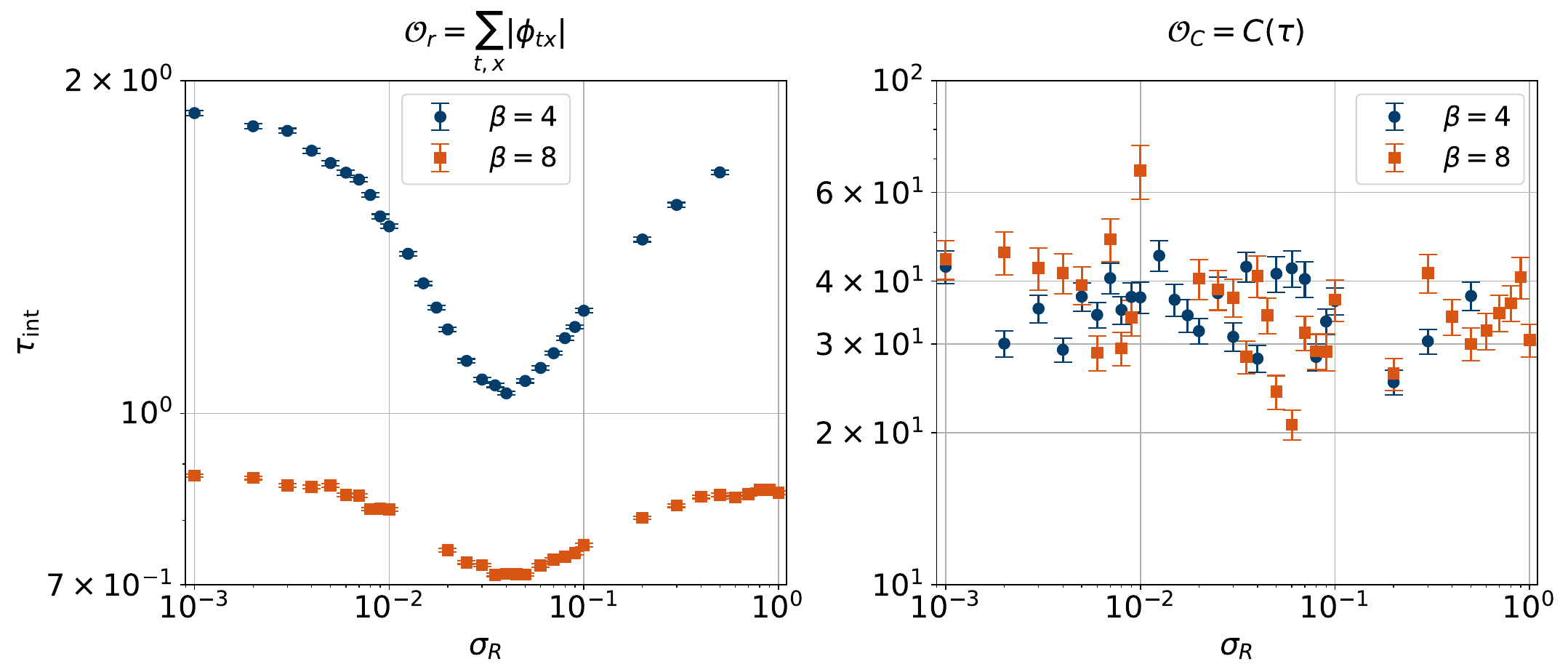}
	\caption{Integrated autocorrelation time $\tau_{\mathrm{int}}$ as a function of the proposal standard deviation $\sigma_R$ for the (left) observable $\mathcal{O}_r$ \eqref{PeRU_obs} and the (right) single particle correlator~\eqref{single particle correlator} projected to the $B^3_1$ operator of~\R{RodekampPerylene}.
		The underlying simulations target the molecule perylene, modeled with the Hubbard model using $N_x=20$, $N_t=96$, $U = 2$, $\kappa = 1$, and $\beta = 4$ (blue circles) or $\beta = 8$ (red squares).}
	\label{fig:PeRU_tint_sig}
\end{figure}

As a final step in our analysis we involve radial updates in a realistic simulation of the molecule $C_{20}H_{12}$ perylene at half-filling. 
The molecule has raised interest due to its applications in organic electronics, such as solar cells \cite{TangOrganicPhotovoltaic,PeryleneFilm,ProgressPeryleneOrganicSolar} and semiconductors \cite{PeryleneElectronTransportLayer,PeryleneOrganicTransistors}, and was recently subject to a Hubbard-model study using HMC in \R{RodekampPerylene}.
In close similarity to this study, we set $\kappa = 1$, $\beta = 4, 8$, $U=2$, and $N_t = 96$.\footnote{For reliable physical insight it is essential to perform the continuum extrapolation $N_t\rightarrow \infty$. For our purposes, it suffices to examine the algorithm's performance at sufficiently large $N_t$.} 
We employ an HMC with trajectory length \eqref{optT} and $N_{\mathrm{MD}} = 6, 10$, resulting in a coarse MD integration with acceptance rates of about $70\%$.
Before every HMC step, we employ a single radial update and we record a total of $N_{\mathrm{conf}} = 2\times 10^6$ configurations, saving after each HMC step.
As in \secref{R2S_tint}, we perform multiple simulations across a range of $\sigma_R$ values to examine the tuning of the radial updates and to assess their impact on autocorrelation times.
Specifically, we compute the integrated autocorrelation times for two representative observables. 
The observables considered are the $L_0$-norm of the field, defined as
\begin{align}
	\label{eq:PeRU_obs}
	\mathcal{O}_r = \sum_{t,x}|\phi_{tx}|,
\end{align}
and the single particle correlator~\eqref{single particle correlator} projected to \eqref{R2S_correlator} the $B^3_1$ operator of \R{RodekampPerylene}.
It is important to note that the autocorrelation function \eqref{autocorr_function} of the correlator exhibits heavy tails, causing the method presented in \R{WolffAutocorr} for selecting the summation window truncating the infinite sum \eqref{tint} to significantly underestimate $\tau_{\mathrm{int},C}$.
To address this, we instead use the point of first zero-crossing of the autocorrelation function to determine the summation window, which proves to be more robust for this particular observable.
As before, for the two-point correlator we select the largest autocorrelation time \eqref{tint_correlator} to reduce the integrated autocorrelation times of the $N_t$ components of the correlator to a single value.
The results are depicted in \figref{PeRU_tint_sig}.
\begin{figure}
	\centering
	\includegraphics[width = 0.9\textwidth]{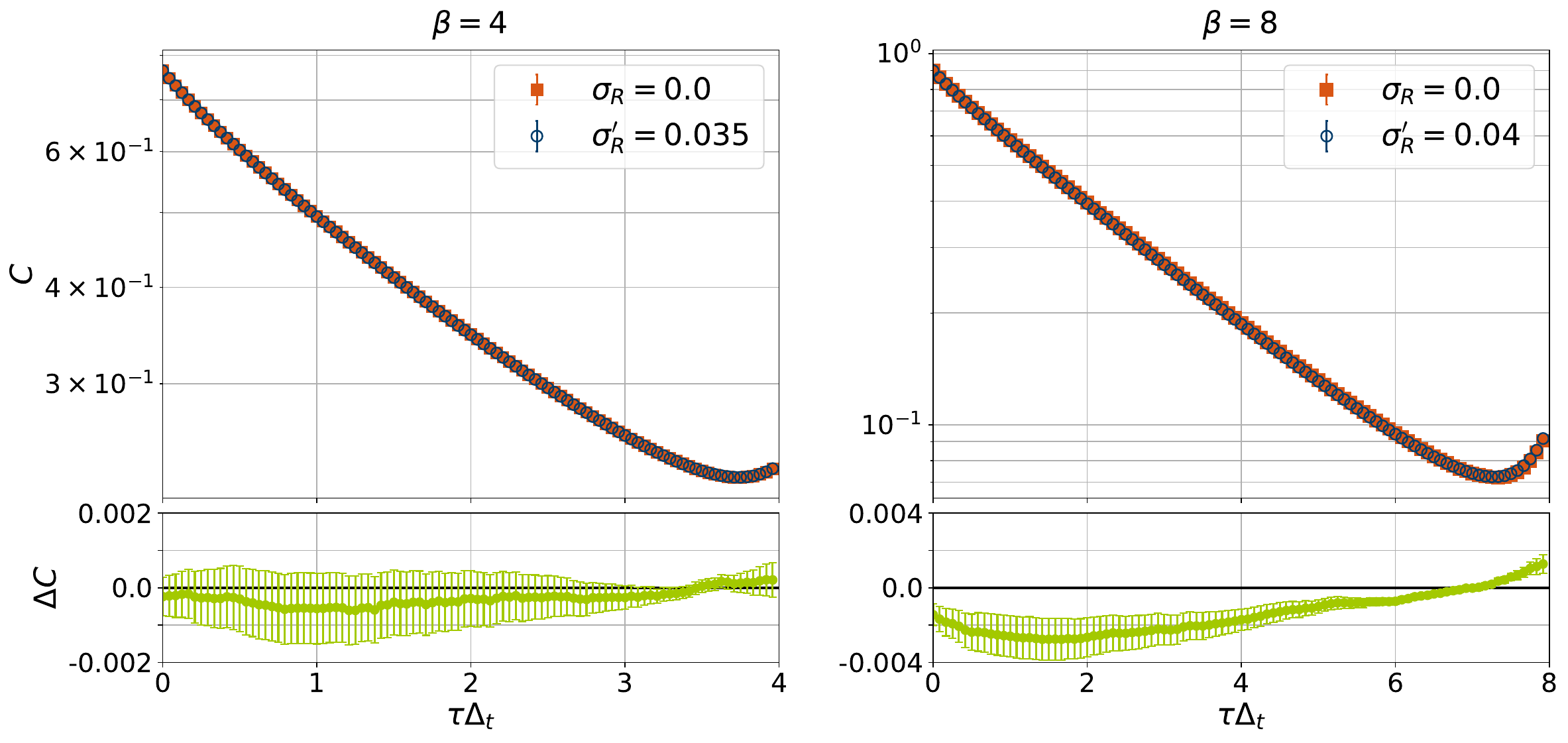}
	\caption{Comparison of the perylene single particle correlators between simulations without radial updates (red squares) and simulations with radial updates using a near-optimal proposal standard deviation (empty blue circles).
		The plots in the bottom margin display the differences between the two correlators.
		Results for $\beta = 4$ are shown on the left, while results for $\beta = 8$ are shown on the right.
	}
	\label{fig:PeRU_corr_comp}
\end{figure}

For the observable $\mathcal{O}_r$, we observe behavior similar to that seen in previous sections for both $\beta = 4$ and $\beta = 8$. 
Specifically, the integrated autocorrelation is maximal near $\sigma_R=0$ and for large $\sigma_R$, with a distinct minimum occurring in between.
This pattern reflects that $\tau_{\mathrm{int}}$ is at its maximum when no radial updates are applied ($\sigma_R=0$), proposed radial updates are very small ($\sigma_R\approx 0$), or the radial acceptance rate is negligibly small ($\sigma_R \gg 0$).
Notably, we find that $\tau_{\mathrm{int}}$ saturates in these outer regions, because a coarse MD integration was used, which allows potential barriers to be crossed regularly.

Given the similarity of the qualitative behavior of $\mathcal{O}_r$ to that observed in the preceding section, we fit \eqref{tint_fit} to estimate the optimal proposal standard deviation $\sigma_R^{\mathrm{(min)}}$.
Fitting the data near the minimum produces the estimates $\sigma_R^{\mathrm{(min)}}=0.03811(25)$ for $\beta = 4$ and $\sigma_R^{\mathrm{(min)}}=0.03907(61)$ for $\beta = 8$.
Comparing this to the value $1/\sqrt{d}\approx 0.023$, obtained from the leading-order scaling in \R{OstmeyerRadialUpdates}, we find that it provides a near-optimal initial guess.
Additionally, we observe a radial acceptance rate of approximately $30\%-35\%$ in the vicinity of $\sigma_R^{\mathrm{(min)}}$, which closely aligns with the optimal range found previously.
The consistent observation of a radial acceptance rate within $30\%-35\%$ throughout \Secref{Results} suggests it can be used as an easily accessible metric for tuning radial updates.
Specifically, we recommend starting with $\sigma_R \approx 1/\sqrt{d}$ and adjusting it to achieve a radial acceptance rate of about $30\%$.
Choosing slightly larger $\sigma_R$ is generally preferable, as it only leads to a linear increase in $\tau_{\mathrm{int}}$, compared to the quadratic scaling observed for smaller $\sigma_R$.

In contrast to the first observable, for the correlator $C(\tau)$, the integrated autocorrelation time exhibits no significant sensitivity to the tuning of radial updates as depicted in the right panel of \figref{PeRU_tint_sig}.
Instead, we observe fluctuating autocorrelation times, indicating that any potential improvement due to radial updates could not be detected at the current level of statistical precision.
Moreover, this could suggest that that the coarse MD integration already renders the simulation ergodic and the radial updates have a negligible effect on reducing autocorrelation times for this particular observable.
This hypothesis is further supported for $\beta = 4$ by comparing the correlator obtained from a standalone HMC simulation ($\sigma_R = 0$) with that from an HMC simulation augmented with radial updates ($\sigma_R'=0.035$).
These correlators are depicted in the left panel of \figref{PeRU_corr_comp}, with their difference, $\Delta C(\tau) = C^{(\sigma_R')}(\tau) - C^{(\sigma_R)}(\tau)$, visualized at the bottom margin.
The results demonstrate that both simulations yield consistent correlators, with any differences falling well within the statistical uncertainties.

At lower temperatures (larger $\beta$), ergodicity violations are expected to become more pronounced, raising the question of whether the observed behavior holds for the $\beta = 8$ simulations.
As shown in \figref{PeRU_tint_sig}, for $\beta = 8$, $\tau_{\mathrm{int}}$ exhibits the same fluctuating behavior as for $\beta = 4$. 
Comparing the correlators obtained from standalone HMC to HMC augmented with radial updates ($\sigma_R'=0.04$), depicted in the right panel of \figref{PeRU_corr_comp}, it is observed that they exhibit the same shape but their difference is only zero within the $3\sigma$ confidence interval.
While this is a statistically significant deviation, it appears small enough to be attributed to statistical fluctuations due to the overall small magnitude of the difference and the large fluctuations in $\tau_{\mathrm{int}}$.
\begin{figure}
	\centering
	\includegraphics[width = 0.8\textwidth]{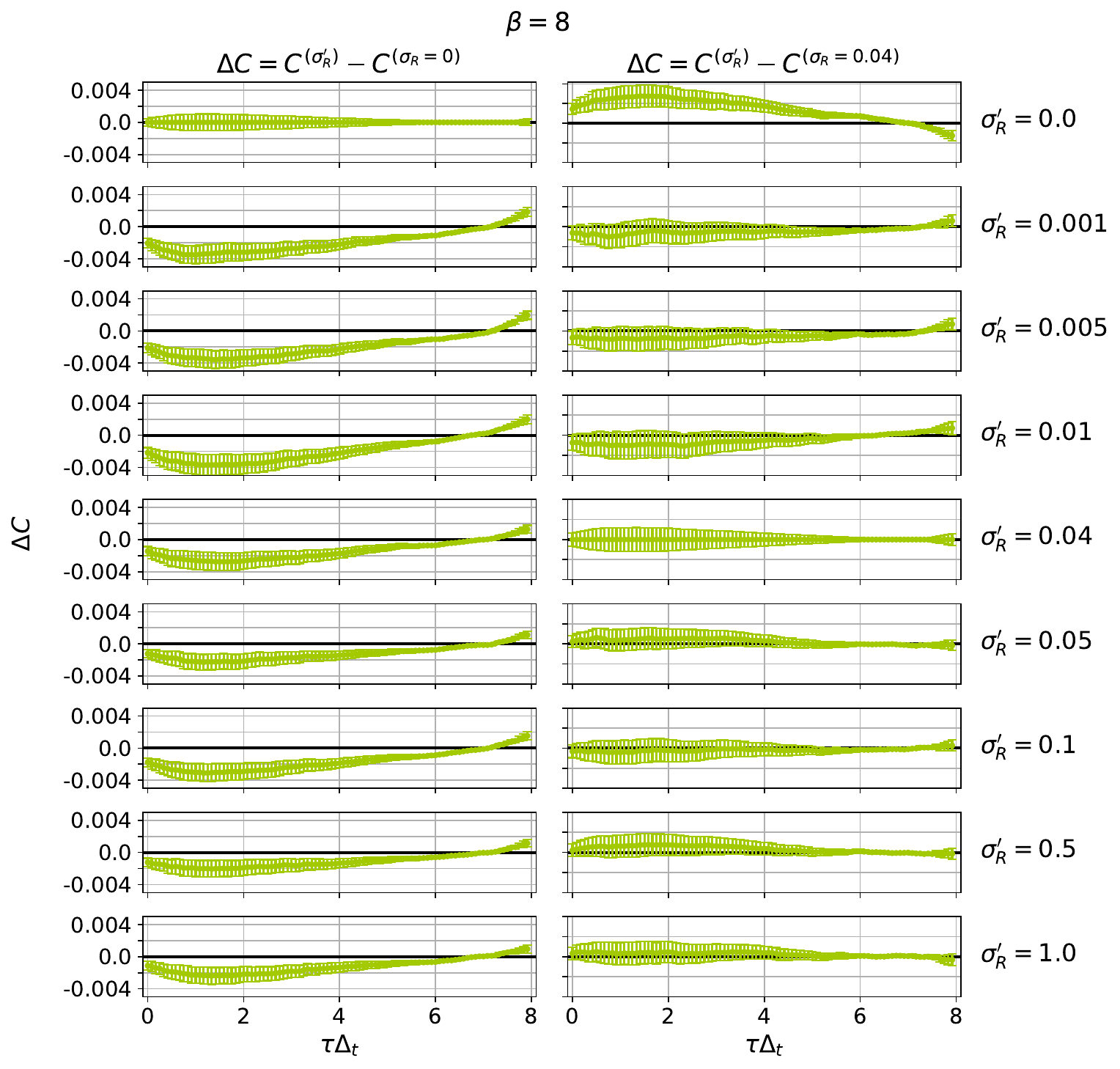}
	\caption{Differences in single particle correlators between simulations with varying proposal standard deviation $\sigma_R'$, defined as $\Delta C = C^{(\sigma_R')} - C^{(\sigma_R)}$.
		The left column compares simulations at varying $\sigma_R'$ to the simulation without radial updates ($\sigma_R= 0$), while the right column compares them to a simulation with near-optimal radial updates ($\sigma_R= 0.04$).
		The simulations target the molecule perylene, modeled with the Hubbard model using $N_x=20$, $N_t=96$, $U = 2$, $\kappa = 1$, and $\beta = 8$.}
	\label{fig:PeRU_corr_comp_b8_full}
\end{figure}
However, a comparison of the differences across all simulations, as visualized in the left column of \figref{PeRU_corr_comp_b8_full} for a selection of simulations, reveals that the deviation is systematic between \emph{all} simulations with radial updates and the simulation without radial updates.
Therefore, although the deviation is small, this indicates that the standalone HMC simulation is biased and suffers from an ergodicity problem.
By contrast, when comparing simulations with radial updates to the close-to-optimal choice $\sigma_R = 0.04$, as visualized in the right column of \figref{PeRU_corr_comp_b8_full}, the differences in correlators fluctuate around zero, as expected for independent and ergodic simulations.
This strongly suggests that the radial updates remove the ergodicity violations, showcasing their effectiveness in simulations of realistic system size.

Finally, it is worth emphasizing that, in the simulation without radial updates, the coarse MD integration repeatedly transitioned through potential barriers and the observed deviations would likely vanish with even coarser MD integration or larger sample size.
However, this approach becomes increasingly computationally expensive, thus further underscoring the necessity of radial updates or alternative techniques to formally address ergodicity violations.

\section{Summary and Outlook}
\label{sec:Summary}
In this work, we proposed to augment the HMC simulation of a lattice field theory defined on a non-compact manifold with a multiplicative Metropolis-Hastings update in the radial direction of the fields.
These so-called radial updates effectively address ergodicity violations caused by infinite potential barriers, and guarantee improved convergence of the algorithm in general \cite{original_radial_update,OstmeyerRadialUpdates}.

Using the Hubbard model as a primary example, we demonstrated that radial updates successfully restore ergodicity by enabling transitions between regions separated by infinitely high potential barriers caused by vanishing fermion determinants.
Furthermore, we examined the scaling behavior of integrated autocorrelation times with the inclusion of radial updates, consistently observing polynomial scaling, which also indicates the absence of ergodicity violations in practice.

Additionally, we analyzed the tuning of the proposal standard deviation $\sigma_R$ introduced by radial updates and confirmed a leading-order scaling or $\sigma_R\propto d^{-0.5}$, where $d$ is the dimensionality of the system.
Empirically, the optimal choice of $\sigma_R$ consistently resulted in a radial acceptance rate within the range of $30\%-35\%$, further offering an accessible and practical guideline for parameter optimization.
We emphasize that, when in doubt, it is preferable to choose a larger $\sigma_R$, as this only leads to a linear increase in $\tau_{\mathrm{int}}$, compared to the quadratic scaling established for smaller $\sigma_R$.

Finally, we incorporated radial updates into a realistic simulation of perylene ($C_{20}H_{12}$), demonstrating that they effectively resolve ergodicity violations and reduce autocorrelation times.

In summary, radial updates offer a computationally efficient and easily tunable method for ensuring ergodicity and convergence of HMC simulations on non-compact manifolds, thereby improving the overall efficiency of the algorithm.
The most computationally expensive aspect of radial updates is computing the change in action, which requires evaluating the fermion determinant.
In this study, we employed exact determinant computations in our simulations; however, incorporating pseudofermions to improve scalability is straightforward.
While this work primarily focused on addressing ergodicity violations in the Hubbard model, similar challenges, such as topological freezing, also arise in lattice gauge theories formulated on compact spaces.
Although directly generalizing radial updates to compact manifolds is not straightforward, extending this approach or augmenting HMC with analogous Metropolis-Hastings schemes presents an intriguing avenue for future research.

\begin{acknowledgments}
	The authors would like to thank Marcel Rodekamp for valuable insights and support in setting up the simulations.
	We also extend our gratitude to Petar Sinilkov for providing exact results for comparison and offering engaging discussions.
	All data used in the analysis of this study are available from the corresponding author upon reasonable request.
	This work was funded in part by the Deutsche
	Forschungsgemeinschaft (DFG, German Research Foundation) as part of the CRC 1639 NuMeriQS–project no.~511713970.
	We gratefully acknowledge the computing time granted by the JARA Vergabegremium and provided on the JARA Partition part of the supercomputer JURECA at Forschungszentrum Jülich~\cite{jureca-2021}.
\end{acknowledgments}


%

\begin{appendix}
	\section{Toy model}
	\label{sec:toy_model}
	In this section, we investigate the impact of radial updates when additionally using coarse MD integration and further analyze possible limitations by again considering the toy model \eqref{toy_model}.
	As discussed in \Secref{toy_model_interlude}, this model is easily solvable, while also exhibiting infinite potential barriers, such that it constitutes a well-controlled testing ground for further assessing the effectiveness of radial updates.
	Moreover, in this model, increasing the dimensionality $d$ may pose a challenge for radial updates, as the probability mass in the central region diminishes with higher $d$ due to fewer possible configurations near the center.
	This poses a potential limitation for radial updates, as effective sample decorrelation may rely on their ability to efficiently jump close to the origin, where HMC can rapidly change the angle in field space.
	Additionally, as $d$ grows, transitions through potential barriers of individual field components are expected to become less frequent, resulting in longer autocorrelation times.
	In the worst-case scenario, simulations may exhibit exponentially increasing autocorrelation times, indicating an in-practice ergodicity problem.
	
	To investigate these dynamics, we focus on two representative observables, specifically, the $L_0$-norm of the entire field and of a single component, i.e.
	\begin{equation}
		\label{eq:toy_model_observables}
		\quad \mathcal{O}_r=\sum_{i=1}^{d}|x_i| \quad \text{and} \quad \mathcal{O}_{r_0}=|x_0|. 	
	\end{equation}
	We closely follow the analysis strategy outlined in \Secref{R2S_tint}, conducting multiple simulations at varying $\sigma_R$ for dimensionalities ranging from $d=1$ to $d=128$ at $\beta = 0.125$.
	These simulations were carried out using both a fine MD integration, achieving acceptance rates of $\gtrsim 99\%$, and coarse MD integration, with acceptance rates of approximately $65\%-75\%$.
	Details on the specific choices of $d$ and simulation parameters are provided in \Secref{sim_details_Rtoy}.
	
	\begin{figure}
		\centering
		\includegraphics[width = 0.9\textwidth]{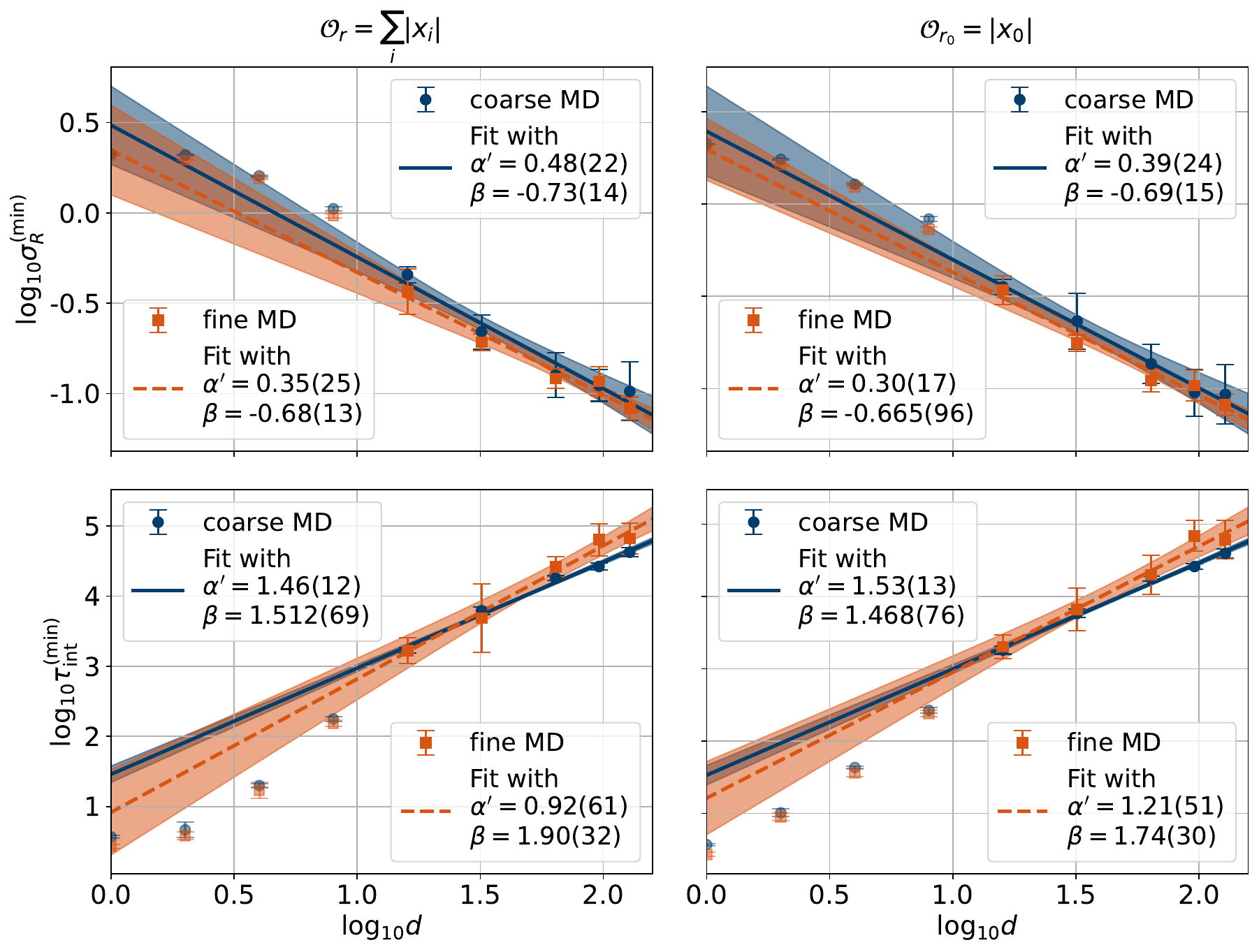}
		\caption{Double logarithmic plots of the optimal proposal standard deviation $\sigma_R^{(\mathrm{min})}$ (top row) and the minimal integrated autocorrelation time  $\tau_{\mathrm{int}}^{(\mathrm{min})}$ (bottom row) as functions of dimensionality $d$.
			The left panels correspond to the field-averaged observable $\mathcal{O}_r$, while the right panels display results for the single-component observable $\mathcal{O}_{r_0}$.
			Simulations were performed using both coarse MD integration (blue circles) and fine MD integration (red squares).
			The data points are fitted to the ansatz \eqref{sigmin_fit} and \eqref{tintmin_fit}, respectively, to determine the leading-order scaling.
			Fit results are provided in the legends, where $\alpha'=\log_{\mathrm{10}}\alpha$.}
		\label{fig:Rtoy_scaling}
	\end{figure}
	Our analysis begins by estimating integrated autocorrelation times \eqref{tint} for the observables \eqref{toy_model_observables} and employing the ansatz \eqref{tint_fit} to estimate the optimal proposal standard deviation $\sigma_R^{(\mathrm{min})}$ as a function of dimensionality $d$. 
	The results, displayed in the top row of \figref{Rtoy_scaling}, indicate that the optimal proposal standard deviation at a given dimensionality is consistent for both observables and compatible between the coarse and fine MD integration approaches.
	To determine the leading-order scaling, we use the fit ansatz \eqref{sigmin_fit}, excluding data points that are shown with low opacity in \figref{Rtoy_scaling} from the fitting procedure. 
	This ensures a more accurate characterization of the leading-order scaling behavior at large $d$.
	Our analysis shows that all the obtained leading-order exponents $\beta$ align with the theoretical $d^{-0.5}$ scaling within the $2\sigma$ confidence interval.
	Furthermore, we again observe a stable radial acceptance rate of $30\%-35\%$ near the optimal proposal standard deviations, closely matching the results discussed in \Secref{R2S_tint}.
	
\begin{figure}
		\centering
		\includegraphics[width = 0.95\textwidth]{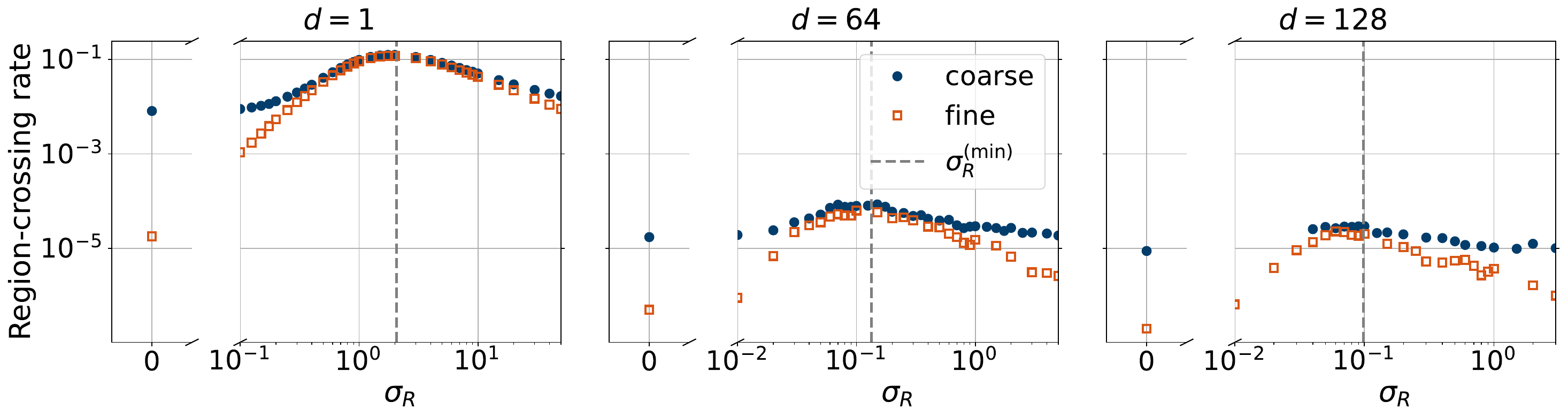}
		\caption{Crossing rates between regions separated by infinite potential barriers as a function of proposal standard deviation $\sigma_R$ used in the radial updates.
			Simulations without radial updates correspond to $\sigma_R= 0$.
			Each panel compares simulations with coarse MD integration (blue circles) to simulations using a fine MD integration (empty red squares).
			The dimensionality increases across panels from left to right, with  $d=1$, $d=64$, and $d=128$, respectively.
		}
		\label{fig:Rtoy_region_crossing}
	\end{figure}

	Next, we determine the integrated autocorrelation time at $\sigma_R^{(\mathrm{min})}$ as a function of dimensionality $d$ and approximate the leading-order scaling using ansatz \eqref{tintmin_fit}.
	The results are depicted in the bottom row of \figref{Rtoy_scaling} and reveal a polynomial scaling for both observables, indicating that any ergodicity violations have been successfully resolved.
	Additionally, we observe slightly longer autocorrelation times with fine MD integration, likely due to the absence of additional tunneling caused by energy violations.
	Furthermore, the scaling between the single component and the field averaged observable aligns for both coarse and fine MD integration, demonstrating that the radial updates consistently facilitate transitions between different regions, even for the single components.
	However, the rate of region-crossings decreases as system size increases, which is illustrated in \figref{Rtoy_region_crossing}.
	Notably, these findings highlight that coarse MD integration itself acts as a non-negligible source of transitions between separated regions in configuration space. 
	Incorporating and fine-tuning radial updates enhances the tunneling rate, thereby improving the overall efficiency of the algorithm.

	\section{Simulation details}
	\label{sec:sim_details}
	In the following we provide details on the parameters used in simulations in the present work. 
	
	\subsection{Toy model}
	\label{sec:sim_details_Rtoy}
	For the toy model, we performed a series of simulations for increasing dimensionality ($d$), utilizing radial updates with varying proposal standard deviations ($\sigma_R$).
	For each choice of $d$, we used a constant HMC trajectory length of $T=1$. 
	The number of MD steps ($N_{\mathrm{MD}}$) was tuned to obtain both a fine MD integration, achieving acceptance rates of $\gtrsim 99\%$, and coarse MD integration, with acceptance rates of approximately $65-75\%$.
	One radial update was applied prior to each HMC step, and measurements were taken after every HMC step.
	To obtain reliable estimates of the integrated autocorrelation times, especially for the single-component observables, the number of recorded configurations ($N_{\mathrm{conf}}$) was increased significantly for dimensions $d>16$.
	The specific choices of $d$, the corresponding $N_{\mathrm{MD}}$ for coarse and fine MD integration, and $N_{\mathrm{conf}}$ used in these simulations are summarized in \tabref{Rtoy_sim_params}.
	\begin{table}[h]
		\centering
		\begin{tabular}{|c|c|c|c|c|c|c|c|c|c|}
			\hline
			$d$ & 1 & 2 & 4 & 8 & 16 & 32 & 64 & 96 & 128 \\ \hline
			$N_{\mathrm{MD}}$ (coarse) & 1 & 2 & 2 & 3 & 4 & 5 & 7 & 8 & 10 \\ \hline
			$N_{\mathrm{MD}}$ (fine) & 8 & 12 & 14 & 20 & 24 & 32 & 44 & 56 & 64 \\ \hline
			$N_{\mathrm{conf}}$& $10^6$ & $10^6$ & $10^6$ & $10^6$ & $10^6$ & $10^7$ & $10^7$ & $10^7$ & $2\times 10^7$ \\ \hline
		\end{tabular}
		\caption{Simulation parameters used to generate data for the toy model with increasing dimensionality ($d$), as discussed in \secref{toy_model}. 
			$N_{\mathrm{MD}}$ represents the number of MD steps per trajectory, once for both coarse and fine MD integration, and $N_{\mathrm{conf}}$ denotes the total number of recorded field configurations.}
		\label{tab:Rtoy_sim_params}
	\end{table}
	
	\subsection{Two-site model}
	\label{sec:sim_details_R2S}
	For the two-site model, we performed a series of simulations for an increasing number of time slices ($N_t$), utilizing radial updates with varying proposal standard deviations ($\sigma_R$).
	For each choice of $N_t$, we employed an optimal HMC trajectory length \eqref{optT} and tuned the number of MD steps ($N_{\mathrm{MD}}$) to obtain an acceptance rate of $>99\%$.
	One radial update was applied prior to each HMC step, and measurements were taken after every HMC step. 
	To obtain reliable estimates of the integrated autocorrelation times, the number of recorded configurations ($N_{\mathrm{conf}}$) was increased incrementally for higher dimensionalities.
	The specific choices of $N_t$, the corresponding $N_{\mathrm{MD}}$, and $N_{\mathrm{conf}}$ used in these simulations are summarized in \tabref{R2S_sim_params}.
	\begin{table}[h]
		\centering
		\begin{tabular}{|c|c|c|c|c|c|c|c|}
			\hline
			$N_t$ & 1 & 4 & 8 & 16 & 24 & 32 & 40 \\ \hline
			$N_{\mathrm{MD}}$ & 60 & 50 & 50 & 40 & 45 & 40 & 45 \\ \hline
			$N_{\mathrm{conf}}$ & $2\times 10^5$ & $2\times 10^5$ & $3\times 10^5$ & $3\times 10^5$ & $3\times 10^5$ & $4\times 10^5$ & $5\times 10^5$ \\ \hline
		\end{tabular}
		\caption{Simulation parameters used to generate data for the two-site model with an increasing number of time slices ($N_t$), as discussed in \secref{R2S_ergodicity} and \secref{R2S_tint}. 
			$N_{\mathrm{MD}}$ represents the number of MD steps per trajectory, and $N_{\mathrm{conf}}$ denotes the total number of recorded field configurations.}
		\label{tab:R2S_sim_params}
	\end{table}

	\section{Four-site model}
	\label{sec:Sq4S}
	As an intermediate step toward incorporating radial updates in the simulation of a realistic system size, we examine the four-site Hubbard model on a $2\times 2$ square lattice.
	In complete analogy to \Secref{R2S_tint}, we performed a series of simulations for an increasing number of time slices ($N_t$), employing radial updates with varying proposal standard deviation $\sigma_R$.
	The choices of $N_t$, the number of MD steps ($N_{\mathrm{MD}}$), and the recorded number of configurations ($N_{\mathrm{conf}}$) are listed in \tabref{Sq4S_sim_params}.
	\begin{table}[h]
		\centering
		\begin{tabular}{|c|c|c|c|c|c|c|c|}
			\hline
			$N_t$ & 1 & 4 & 8 & 16 &  32 & 40 \\ \hline
			$N_{\mathrm{MD}}$ & 60 & 60 & 50 & 50 & 50 & 55 \\ \hline
			$N_{\mathrm{conf}}$ & $3\times 10^5$ & $3\times 10^5$ & $3\times 10^5$ & $3\times 10^5$ & $4\times 10^5$ & $4\times 10^5$ \\ \hline
		\end{tabular}
		\caption{Simulation parameters used to generate data for the four-site model on a $2\times 2$ square lattice with an increasing number of time slices ($N_t$). 
			$N_{\mathrm{MD}}$ represents the number of MD steps per trajectory, and $N_{\mathrm{conf}}$ denotes the total number of recorded field configurations.}
		\label{tab:Sq4S_sim_params}
	\end{table}
	
\begin{figure}
		\centering
		\includegraphics[width = 0.8\textwidth]{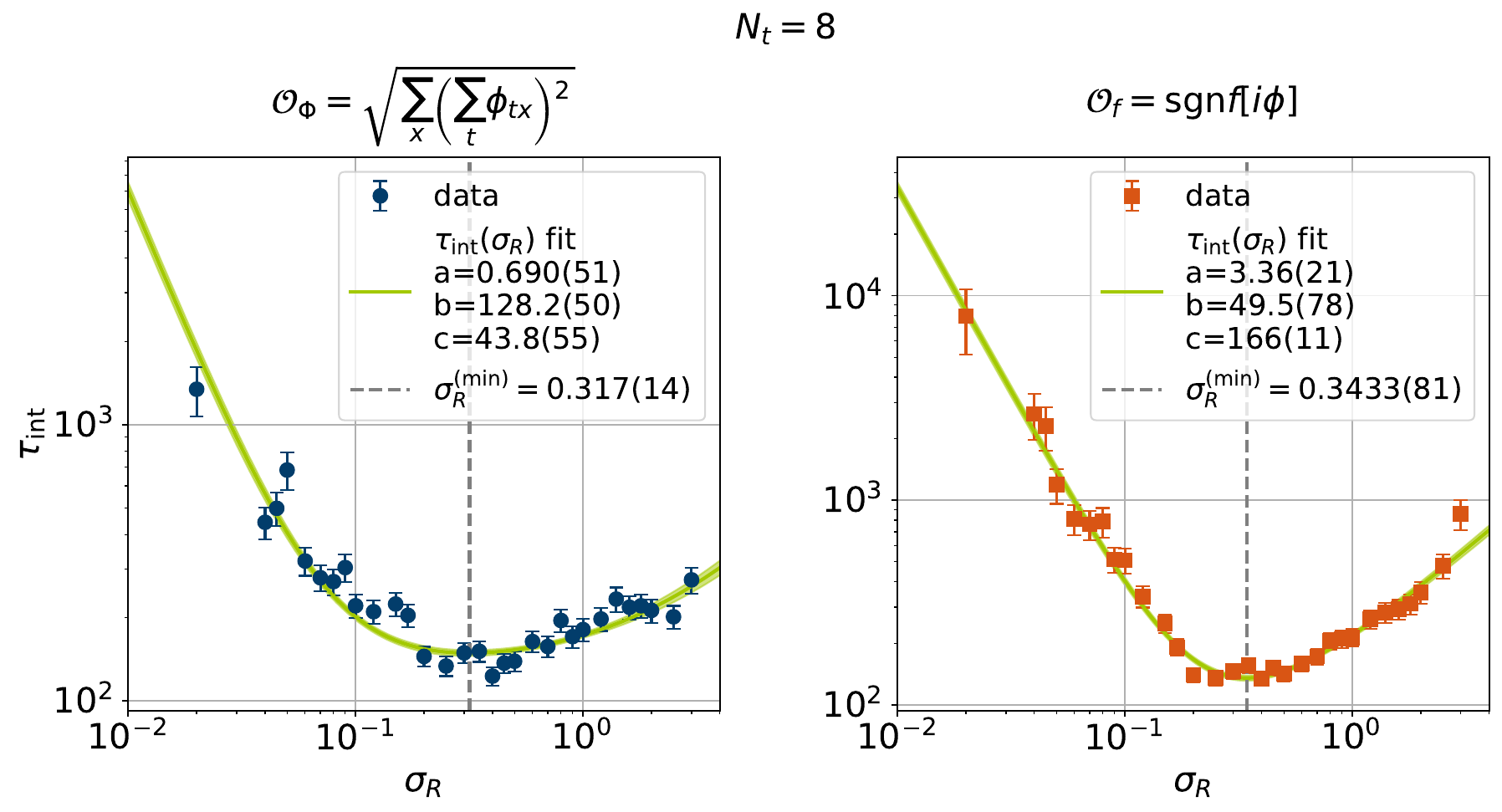}
		\caption{Integrated autocorrelation time $\tau_{\mathrm{int}}$ as a function of the proposal standard deviation $\sigma_R$ for the observables $\mathcal{O}_\Phi$ (left) and $\mathcal{O}_f$ (right). 
			The observables are defined in \eqref{Phi_radius} and \eqref{sgndetM}, respectively.
			The fits (green line) are obtained using the ansatz \eqref{tint_fit}, with fit results shown in the corresponding legends.
			These results are used to estimate the optimal proposal standard deviation $\sigma_R^{\mathrm{(min)}}$ (grey dashed line), whose value is also provided in the legends.
			The underlying model is the four-site Hubbard model on $2\times 2$ square lattice with $N_t=8$ for $U = 18$, $\beta = 1$, and $\kappa = 1$.
			The HMC simulations were conducted using \eqref{optT} and $N_{\mathrm{MD}}=50$, achieving an acceptance rate $>99\%$.
			Each HMC step was preceded by a single radial update and a total of $N_{\mathrm{conf}}=3\times 10^5$ configurations were recorded, with measurements taken after each HMC step.}
		\label{fig:Sq4S_Nt8_tint}
	\end{figure}

\begin{figure}
		\centering
		\includegraphics[width = 0.8\textwidth]{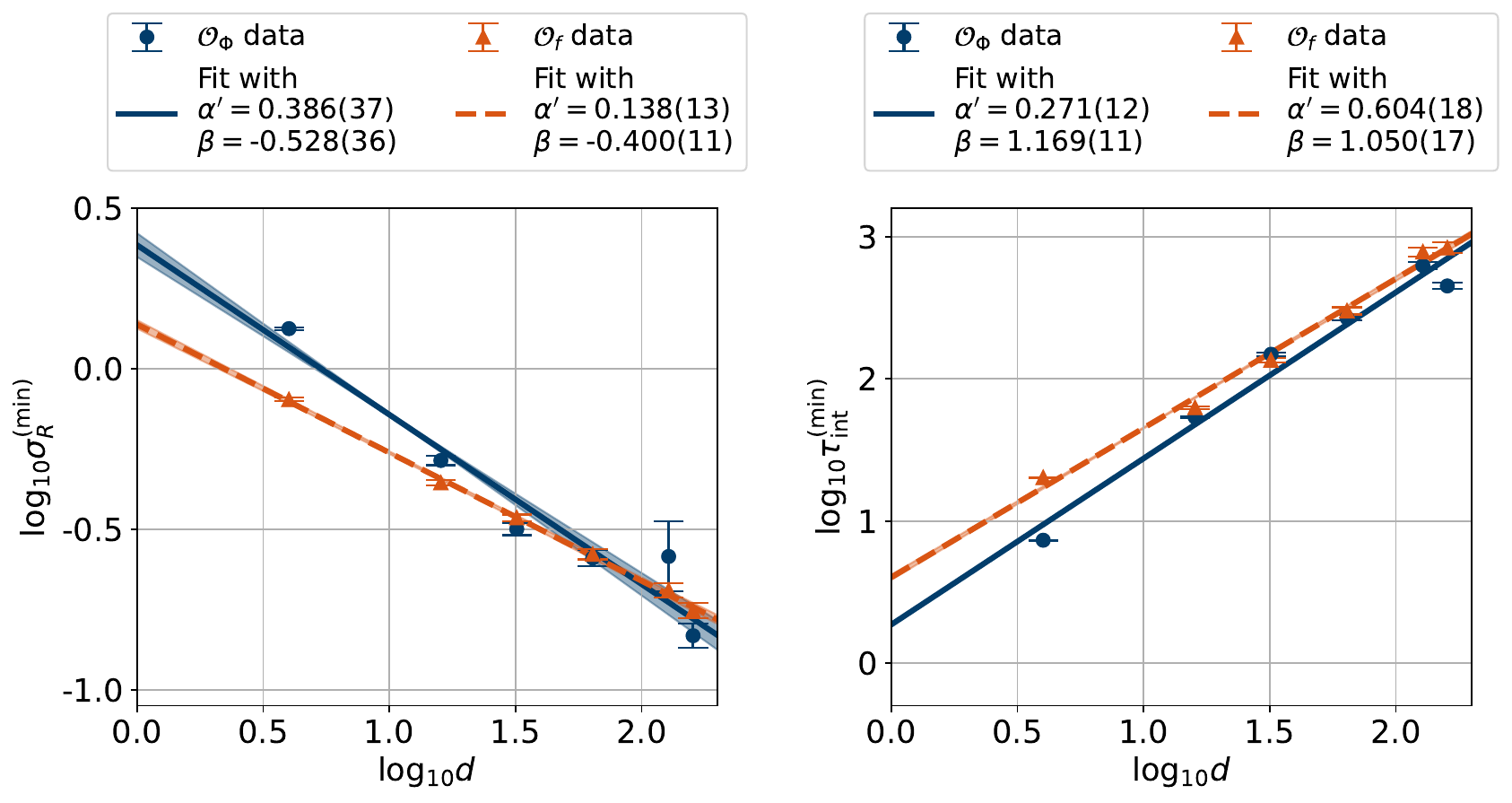}
		\caption{The left panel shows a double logarithmic plot of the position of the minimum $\sigma_R^{(\mathrm{min})}$ as a function of dimensionality $d=N_xN_t$.
			The right panel illustrates a double logarithmic plot of the integrated autocorrelation time at the respective minimum, denoted by $\tau_{\mathrm{int}}^{(\mathrm{min})}$, also as a function of the dimensionality $d$.
			The data in both panels is fitted to the respective ansatz, \eqref{sigmin_fit} and \eqref{tintmin_fit}, to determine the leading-order scaling.
			Fit results are provided in the legends, where $\alpha'=\log_{\mathrm{10}}\alpha$.
			The underlying model is the four-site Hubbard model on a $2\times 2$ square lattice with varying $N_t$ for $U = 18$, $\beta = 1$, and $\kappa = 1$.
		}
		\label{fig:Sq4S_scaling}
	\end{figure}

	Similarly to the two-site model, we compute the integrated autocorrelation time for two representative observables.
	Specifically, we consider the radial observable \eqref{Phi_radius} and the observable
	\begin{align}
		\label{eq:sgndetM}
		\mathcal{O}_f = \mathrm{sgn} f[i\phi|\kappa], \quad \text{where} \quad f[i\phi|\kappa] = \det \left(M[\phi|\kappa]\right)e^{-i\sum_{tx}\phi_{tx}},
	\end{align}
	which first projects the fermion determinant onto a real number and then determines its sign.
	Therefore, this observable is directly related to the infinite potential barriers discussed in \Secref{Hubbard}, which occur when the fermion determinant vanishes, i.e.\@ potentially changes sign.
	For example, the diagonal bands visualized in \figref{Nt1_scatter} exhibit alternating signs of $\mathcal{O}_f$, highlighting the observable's sensitivity to transitions through the potential barriers.
	Examplary results for the simulations at $N_t=8$ are shown in \figref{Sq4S_Nt8_tint}.
	These results confirm that employing and tuning the radial updates significantly reduces the autocorrelation times, particularly for the observable $\mathcal{O}_f$.
	Furthermore, we also identify a distinct minimum in $\tau_{\mathrm{int}}$ and estimate its position by performing fits to the ansatz \eqref{tint_fit}.
	Extending this analysis to all values of $N_t$, we verify the leading-order scaling of the optimal proposal standard deviation, $\sigma_R^{(\mathrm{min})}(d) \propto d^{-0.5}$, by using the fit ansatz \eqref{sigmin_fit}.
	The results, presented in the left panel of \figref{Sq4S_scaling}, demonstrate excellent agreement with the theoretical prediction for the observable $\mathcal{O}_\Phi$.
	For $\mathcal{O}_f$, the deviation from the $d^{-0.5}$ scaling are slightly larger but remain reasonable, given that the predicted scaling is at leading order and fits include data from small $d$ values.
	
	Finally, we determine the minimal integrated autocorrelation time $\tau_{\mathrm{int}}^{(\mathrm{min})}$, measured at the respective optimal proposal standard deviation.
	The results, along with leading-order fits to the ansatz \eqref{tintmin_fit}, are displayed in the right panel of \figref{Sq4S_scaling}.
	We observe a slightly greater than linear scaling for both observables, indicating that radial updates have resolved any ergodicity violations.
	
\end{appendix}

\end{document}